\documentclass[a4paper,10pt,aps,pra,twocolumn,showpacs,superscriptaddress,floatfix]{revtex4-1}
\pdfoutput=1
\usepackage{graphicx}
\usepackage{amssymb}
\usepackage{amsmath}
\usepackage{bbold}
\usepackage{color}

\begin{document}

\title{Reservoir engineering and dynamical phase transitions in optomechanical arrays}

\author{A. Tomadin}
\affiliation{Institute for Quantum Optics and Quantum Information of the Austrian Academy of Sciences, A-6020 Innsbruck, Austria}
\affiliation{NEST, Istituto Nanoscienze-CNR and Scuola Normale Superiore, I-56126 Pisa, Italy}
\author{S. Diehl}
\affiliation{Institute for Theoretical Physics, University of Innsbruck, Technikerstr.~25, A-6020 Innsbruck, Austria}
\author{M.D. Lukin}
\affiliation{Department of Physics, Harvard University, Cambridge MA, USA}
\author{P. Rabl}
\affiliation{Institute for Quantum Optics and Quantum Information of the Austrian Academy of Sciences, A-6020 Innsbruck, Austria}
\author{P. Zoller}
\affiliation{Institute for Theoretical Physics, University of Innsbruck, Technikerstr.~25, A-6020 Innsbruck, Austria}
\affiliation{Institute for Quantum Optics and Quantum Information of the Austrian Academy of Sciences, A-6020 Innsbruck, Austria}

\begin{abstract}
We study the driven-dissipative dynamics of photons interacting with an array of micromechanical membranes in an optical cavity.
Periodic membrane driving and phonon creation result in an effective photon-number conserving non-unitary dynamics, which features a steady state with long-range photonic coherence.
If the leakage of photons out of the cavity is counteracted by incoherent driving of the photonic modes, we show that the system undergoes a dynamical phase transition to the state with long-range coherence.
A minimal system, composed of two micromechanical membranes in a cavity, is studied in detail, and it is shown to be a realistic setup where the key processes of the driven-dissipative dynamics can be seen.
\end{abstract}


\date{\today}

\maketitle

\section{Introduction}
\label{sec:introduction}

While quantum computing and quantum simulation are traditionally discussed as dynamics of isolated many-body systems governed by a unitary time-evolution~\cite{simulator1,simulator2,simulator3,simulator4}, recent interest has turned to engineering and controlling the time-evolution of {\em open} quantum systems.
There, the goal is to design couplings to an environment, so that the many-body time-evolution corresponds to a specified Kraus map or, in the Markovian limit, a master equation (ME)~\cite{MDPZ2012}.
In Ref.~\cite{DMKKBZ2008} a scenario of open system dynamics was discussed, where driven-dissipative dynamics drives the system of interest into a (desired) entangled steady state, and where competition between the coherent Hamiltonian and the dissipative Liouvillian terms in the ME gives rise to a rich nonequilibrium phase diagram and dynamical phase transitions~\cite{DTMFZ2010,TDZ2011}, which do not have immediate condensed matter counterparts.
Such far-from-equilibrium systems may thus open up new perspectives for many-body physics, challenging both theory and experiment.
In addition, engineered dissipation also provides the basis for a dissipative variant of quantum computing, as first suggested in Ref.~\cite{VWC2009}.

Implementing these concepts of reservoir engineering of quantum many-body systems has been so far mainly discussed with cold atoms and ions, and first experiments have demonstrated driven-dissipative preparation of a four-particle Greenberger-Horne-Zeilinger (GHZ) states with ions~\cite{BMSNMCHRZB2011}, and Einstein-Podolsky-Rosen (EPR) states of atomic ensembles~\cite{KMJWPCP2011}.
In light of the recent discussion of cavity arrays and photonic quantum simulation~\cite{HBP2008,TF2010}, it seems natural to ask to what extent the concepts of quantum reservoir engineering and associated nonequilibrium phenomena can be realized with coupled photonic systems.
In cavity arrays, photon loss appears as a natural decoherence mechanism and a steady state can only be sustained under nonequilibrium conditions, i.e.~by continuously driving the system with external laser fields~\cite{H2010,NSBBTK2012}.
However, the design of non-trivial couplings of optical photons to an engineered reservoir, leading, for example, to number-conserving or non-local dissipation processes, is harder to achieve and imposes a challenge for these systems. 
In view of the remarkable progress in the field of opto-nanomechanics, we will be interested below in setups where the nanomechanical elements play the role of quantum reservoirs, and a properly designed coupling of nanomechanical oscillators to cavity modes results in the desired quantum dissipation. 

In conventional optomechanical systems (OMS)~\cite{KV2008,MG2009} a single optical mode is coupled via radiation pressure interactions to the motion of a macroscopic mechanical resonator, represented, for example, by a moving end-mirror or a vibrating membrane~\cite{TZJMGH2008} inside a Fabry-Perot resonator.  
Over the past years this coupling has been successfully employed to cool mechanical systems close to the quantum ground state~\cite{Groeblacher2008,VDWSK2012, CMSHKGAP2011,RNMHCS2010,TDLHACSWLS2011}, using  techniques analogous to laser cooling of atoms.
In parallel, rapid progress in the fabrication and control of OMS, and in particular new designs for micro- and nano- scale devices~\cite{VDWSK2012,CMSHKGAP2011,ECCVP2009,DBSLDLF2010}, have led to a drastic improvement of OMS and pave the way for realizing various strongly coupled~\cite{LudwigNJP2008,RablPRL2011,NunnenkampPRL2011} and multi-mode~\cite{GrudininPRL2010,DobrindtPRL2010,MiaoPRL2009,LSPM2012,SKHBLZR2012,BM2008,XGD2012} scenarios.
Here we describe the appearance of novel dissipation processes in extended OM arrays, where in contrast to OM laser cooling, now the mechanical systems provide a decoherence channel for light.
In particular, this mechanism already provides a basic building block for reservoir engineering for light, and we discuss how this can be used to study the dissipation-induced preparation of photonic states with long-range coherence~\cite{MTDR2012}. 

The paper is organized as follows.
In Sec.~\ref{sec:introsystems} we review the key concepts behind the engineering of driven-dissipative quantum many-body systems.
In particular we introduce the driven-dissipative Bose-Hubbard dynamics discussed in Ref.~\cite{DMKKBZ2008} for cold atoms, which yields a Bose-Einstein condensate (BEC) in steady state.
In Secs.~\ref{sec:twomembranes} and \ref{sec:membranesarray} we show how the analogous ME for driven-dissipative dynamics of photons in an optomechanical (OM) system can be engineered. 
In Sec.~\ref{sec:phasetransition} we show that, under continuous incoherent pumping, the system undergoes a nonequilibrium phase transition.
Finally, in Sec.~\ref{sec:conclusions} we summarize our findings.

\section{Driven-dissipative quantum many-body systems}
\label{sec:introsystems}

In this section we briefly introduce and review the many-body ME for driven-dissipative dynamics.
The specific example is dissipative Bose-Hubbard dynamics, as discussed previously for cold atoms in optical lattices immersed in an atomic BEC, which plays the role of a reservoir of phonons.
The dissipative dynamics drives the system in steady state into a BEC, representing a ``dark many-body state.''
This discussion will set the stage for the MEs for photons interacting with an array of micromechanical membranes in an optical cavity in the following sections.

\subsection{Engineered steady states}

The time-evolution of a quantum system that is weakly coupled to an energy or particle reservoir is governed, in the Markov approximation, by the ME
\begin{equation}\label{eq:masterequation}
\partial_{t} \rho(t) = -i [\hat{\cal H}, \rho(t)] + {\cal L}[\rho(t)]~,
\end{equation}
where $\rho(t)$ is the density matrix of the system, $\hat{\cal H}$ the Hamiltonian generating unitary dynamics, and the Liouvillian ${\cal L}$ is a non-unitary, dissipative contribution.
In general, the Liouvillian ${\cal L}$ can be written as a sum $\sum_{\ell} \kappa_{\ell} \Lambda[\hat{K}_{\ell}][\rho(t)]$ of terms in Lindblad form
\begin{equation}\label{eq:genericlindblad}
\Lambda[\hat{K}_{\ell}][\rho(t)] = 2 \hat{K}_{\ell} \rho(t) \hat{K}_{\ell}^{\dag} - \hat{K}_{\ell}^{\dag} \hat{K}_{\ell} \rho(t) - \rho(t) \hat{K}_{\ell}^{\dag} \hat{K}_{\ell}~,
\end{equation}
where $\{ \hat{K}_{\ell} \}_{\ell}$ is a set of so-called jump operators.
In the case of a photonic system on a lattice, for example, with annihilation operator $\hat{a}_{\ell}$ on the $\ell$th site, local losses of photons are described by $\hat{K}_{\ell} = \hat{a}_{\ell}$.
The non-unitary part of the evolution competes with the interaction Hamiltonian, e.g.~the local Kerr nonlinearity $\hat{\cal H} = \frac{U}{2} \sum_{\ell} \hat{a}_{\ell}^{\dag} \hat{a}_{\ell}^{\dag} \hat{a}_{\ell} \hat{a}_{\ell}$, and in general the steady state of the system cannot be fixed by controlling the eigenstates of $\hat{\cal H}$ only.
Under the action of these local jump operators, a pure initial many-body state $|\Psi\rangle$, extended over several lattice sites, evolves into a mixed state.

Recently, jump operators of a novel kind have been designed, which drive the system towards a well-defined, pure many-body steady state
\begin{equation}\label{eq:purity}
\rho(t) \xrightarrow{t\to\infty} |\Psi \rangle \langle \Psi |~.
\end{equation}
The engineering of such jump operators is based on the combination of coherent driving and dissipation processes, hence the resulting models are properly dubbed driven-dissipative systems.
The main goal of such reservoir engineering is to obtain many-body states $|\Psi\rangle$ which feature interesting properties, e.g.~long-range or topological order.
These target states do not necessarily have to be associated to the ground state of some physical Hamiltonian and do not result from minimizing energy, but are stabilized by a combination of drive and dissipation.

The case in which $|\Psi\rangle$ represents an atomic BEC of neutral atoms in an optical lattice has been considered in Ref.~\cite{DMKKBZ2008}.
In the condensed state
\begin{equation}\label{eq:becstate}
|{\rm BEC}\rangle = \frac{1}{\sqrt{N!}} \left ( \frac{1}{\sqrt{L}} \sum_{\ell = 1}^{L} \hat{a}_{\ell}^{\dag} \right )^{N} |{\rm vac}\rangle~,
\end{equation}
all $N$ bosonic atoms occupy the same single-particle state, given by the symmetric superposition of the $L$ lattice sites.
In the absence of interactions, the target state $|{\rm BEC}\rangle$ is obtained by the bilocal jump operator
\begin{equation}\label{eq:paradigmaticjump}
\hat{K}_{\ell} \propto (\hat{a}_{\ell+1}^{\dag} + \hat{a}_{\ell}^{\dag}) (\hat{a}_{\ell+1} - \hat{a}_{\ell})~.
\end{equation}
A weak interaction, taken into account perturbatively, acts on the steady state as an effective temperature $k_{\rm B}T \propto U$ and spoils the purity of the final state.
When the interactions are so strong that they dominate the dynamics, the steady state is mixed and described by a diagonal density matrix.
In the intermediate regime where the two energy scales compete, it has been found that a phase transition takes place \cite{DTMFZ2010,TDZ2011}, which shares features of a quantum phase transition in that it is driven by the competition of two non-commuting operators, and of a statistical phase transition in that the ordered phase directly terminates into a strongly mixed state.

In the case of a photonic system, it is necessary to complement the Hamiltonian dynamics with a nonunitary term ${\cal L}$ to take into account the decay of the photons.
More precisely, the lifetime of the photons is usually shorter than the timescale $\propto U^{-1}$ over which the unitary many-body dynamics takes place.
In this case, it is not possible to assume that the photons thermalize, or reach the ground state of the Hamiltonian $\hat{\cal H}$, but the complete dynamics of the system has to be considered, in the presence of pumping and losses.
In general, the existence of a relation between the ground-state properties of a closed many-body system and its open, driven counterpart is not obvious, although it may emerge under specifically tailored drivings~\cite{TGFGCTI2010,NSBBTK2012}.

To design a toolbox for the simulation of driven-dissipative systems would allow to study the relation between the steady-state properties of a nonequilibrium model (\ref{eq:masterequation}) with number-conserving jump operators, e.g.~Eq.~(\ref{eq:paradigmaticjump}), and the corresponding open system with particle losses.
Moreover, the relation between equilibrium and nonequilibrium phases of a model~\cite{KGIYLC2012} could be investigated systematically.
More generally, the flexibility in the implementation of the jump operators could be leveraged to study the effects of the competition between different terms in the nonunitary contribution ${\cal L}$ to the Liouvillian in a driven-dissipative nonequilibrium scenario with stable stationary states, and no immediate condensed matter counterpart.

\begin{figure}
\includegraphics[width=\linewidth]{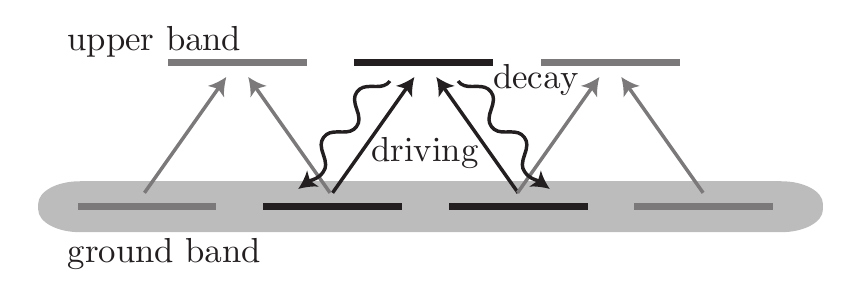}
\caption{\label{fig:modelsketch}
Bosonic modes in two bands of a superlattice.
Coherent driving of neighboring sites from the ground to the upper band and incoherent decay is represented by straight and wavy arrows, respectively.
The whole array supports a chain of $\Lambda$-systems, each defined by a triplet of neighboring modes, two of which belong to the ground band.
Phase locking takes place in the shaded area as a result of the combined action of the chain of $\Lambda$-systems. }
\end{figure}

\subsection{Dark states and $\Lambda$-systems}

The controlled achievement of a given target steady state via a certain driven-dissipative scheme is guaranteed in a situation where the stationary state of the dynamics (\ref{eq:masterequation}) coincides with a unique \emph{dark state} of the Liouvillian, i.e.~for $|\Psi\rangle$ such that $\hat{K}_\ell |\Psi\rangle = 0$ for all $\ell$ \cite{Kraus08,VWC2009}.
For example, it was shown in Ref.~\cite{DMKKBZ2008} that the Bose-Einstein condensate is a dark state of the jump operator (\ref{eq:paradigmaticjump}).
It can be shown to be unique, and moreover to be the unique stationary state of the dissipative dynamics (no stationary states other then the dark state exist in this case) \cite{Kraus08}.
The design of jump operators that feature a given dark state is thus of central importance to the success of a driven-dissipative scheme.
Such task has been accomplished to generate orbital pairing~\cite{DYDZ2010} as well as topological states~\cite{DRBZ2011} in the steady state.

The design of appropriate jump operators is much simplified when it can be reduced to the generation of quasilocal couplings, which are then repeated periodically in the array.
For example, the jump operators (\ref{eq:paradigmaticjump}) act pairwise on the lattice sites, annihilating  any antisymmetric component in the wavefunction on a pair of sites, and mapping it back into the symmetric one.
The repetition of this quasilocal action results in the phase-locking of the wavefunction in the whole array, i.e.~the generation of the long-range order.

The jump operator (\ref{eq:paradigmaticjump}) contains the effective dynamics of a $\Lambda$-system coupled to a dissipative environment, where both the latter and the excited state have been adiabatically eliminated.
Fig.~\ref{fig:modelsketch} illustrates the basic process in which bosons in the ground band are coherently driven (straight lines) with opposite Rabi frequencies to an upper band of excited states.
Only the antisymmetric superposition between neighboring ground modes is excited because of the choice of the Rabi frequencies.
The excited state decays back to a combination of ground states. 
The spontaneous decay is made possible energetically by the creation of an excitation in the reservoir, which carries away the released energy (wavy lines). 
Note that, while energy is exchanged between system and reservoir in this way, the system particle number is conserved during this process---the system constituents merely make transitions between different bands.
A description of these processes in terms of the jump operators (\ref{eq:paradigmaticjump}) is valid given two approximations hold: (i) the Born-Markov approximation underlying the ME description requires the separation of the two bands exceeding all other energy scales in the problem; (ii) the adiabatic elimination of the upper band giving rise to an effective single-band description is justified for a driving process in a regime where the detuning exceeds the Rabi frequencies, as well as the additional energy scales involved in the dynamics, in particular the many-body couplings that arise when several $\Lambda$-systems are joined together. 
We refer to Ref. \cite{MDPZ2012} for a review of the physics of engineered driven-dissipative systems.

In Sec.~\ref{sec:twomembranes} we propose an implementation of the $\Lambda$-system using two micromechanical membranes in an optical cavity, and in Sec.~\ref{sec:membranesarray} we demonstrate that one obtains a set of jump operators analogous to (\ref{eq:paradigmaticjump}), once the adiabatic elimination has been performed.

\section{Optomechanical $\Lambda$-system for photons}
\label{sec:twomembranes}

The implementation of non-trivial dissipative processes as, for example, given by Eq.~(\ref{eq:paradigmaticjump}), has previously been described  in the context of cold atoms by engineering appropriate couplings to a bath of Bogoliubov excitations~\cite{DMKKBZ2008}.
While in photonic systems dissipation arises more naturally, it usually appears in the form of single-photon decay, and the design of both \emph{nonlinear} as well as \emph{number conserving} photon processes imposes a challenge.
In the following, we describe how this can be achieved in OM arrays,  by making use of nonlinear radiation pressure interactions between a set of optical and a set of damped mechanical modes. 

The basic idea is illustrated in Fig.~\ref{fig:lambdasystem}a for a ``membrane in the middle''  setting~\cite{TZJMGH2008}, where (in a minimal version) two semi-transparent membranes separate the optical field into two modes $\hat a_{1,2}$ with frequency $\omega$ and a third mode $\hat c$ with a slightly higher frequency $\omega^\prime$.  Oscillations of the membranes around their equilibrium positions then lead to phonon-assisted scattering of photons between the modes $\hat a_{1,2}$ and $\hat c$, whenever the mechanical vibration frequency $\Omega_{\rm M}$ is close to the optical frequency difference, $\Omega_{\rm M}\approx \omega^\prime-\omega$.
When the mechanical systems are classically driven this scattering is coherent and---in analogy to atomic three level systems---by choosing appropriate phases of the driving fields we can select certain ``bright" and ``dark" combinations of $\hat a_1$ and $\hat a_2$. 
In addition, the coupling of the mechanical systems to a bath of phonons in the support of the membranes, analogously to the coupling to a low temperature photon bath for laser cooling \cite{aspect88,kasevich92},  provides a natural dissipation channel.
Phonon-assisted scattering events followed by a phonon decay become irreversible, and thereby introduce an effective number conserving dissipation process for photons.  

In the remainder of this section we describe in more detail how the combination of coherent and incoherent processes in OM systems can be used to implement a driven-dissipative many-body system for photons.
For this analysis we will focus for concreteness on the ``membrane in the middle'' configuration shown in Fig.~\ref{fig:lambdasystem}a and consider the extension of this model to a whole OM array in Sec.~\ref{sec:membranesarray}.
However, our results are not restricted to this specific setting and apply to various other OM systems.
For example, in panels (c) and (d) of Fig.~\ref{fig:lambdasystem}, we illustrate two equivalent setups, where the OM array is implemented using an array of coupled microtoroidal cavities or coupled OM crystal cavities (see e.g.~Ref.~\cite{SRAAK2008} and Ref.~\cite{ECCVP2009}, respectively).
Due to the strong confinement of the optical field on a few micrometer scale, these setups enable much stronger OM interactions and can be more easily scaled to larger arrays.

\subsection{The model}
\label{ssec:themodel}

\begin{figure}
\includegraphics[width=\linewidth]{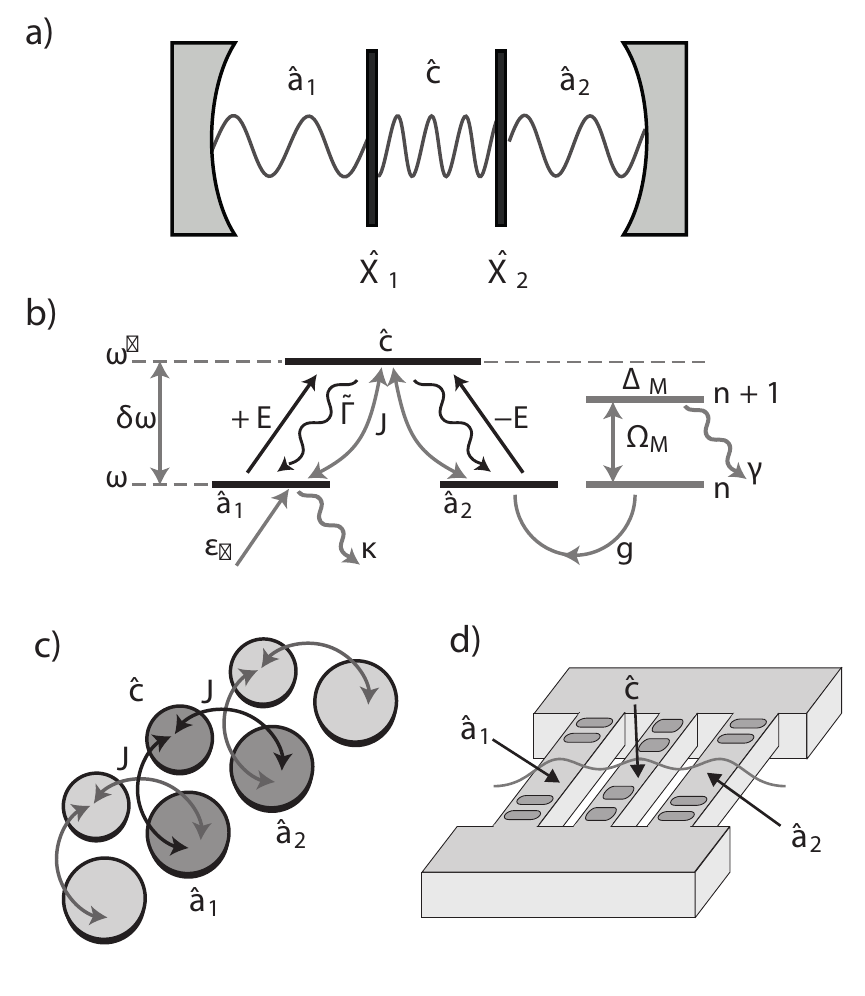}
\caption{\label{fig:lambdasystem}
a) OM setup for implementing engineered dissipation processes for photons. Three optical modes $\hat a_{1,2}$ and $\hat c$ are delimited by two membranes (vertical thick lines) at positions $X_{1}$ and $X_{2}$ and the two fixed end mirrors of the optical cavity.
The modes are tunnel-coupled with amplitude $J$ and their frequencies are modulated by oscillations of the membranes with frequency $\Omega_{\rm M}$ around their equilibrium positions.     
b) The resulting energy level diagram assuming degenerate modes  $\hat a_1$ and $ \hat a_2$ and a slightly higher frequency $\omega^\prime \approx  \omega +\Omega_{\rm M}$ for the intermediate mode $\hat c$ (see text for more details).
c) and d) Alternative implementations using an array of coupled microtoriodal cavities or coupled OM crystal cavities, respectively.
In both cases the optical fields are confined inside the structures, coupled among each other via their evanescent fields and interact with localized mechanical vibrations of the structures via optical gradient forces. }
\end{figure}

We start out by describing the OM setup that we envision.
We consider two micromechanical membranes placed in an optical cavity composed of two macroscopic, high-finesse mirrors (see Fig.~\ref{fig:lambdasystem}a).
The Hamiltonian of the system is of the form
\begin{equation}
\hat{\cal H} = \hat{\cal H}_{\rm M} + \hat{\cal H}_{\rm ph} + \hat{\cal H}_{\rm OM}~,
\end{equation}
where $\hat{\cal H}_{\rm M}$ describes the quantized motion of one vibrational mode of the membranes, $\hat{\cal H}_{\rm ph}$ describes the photonic degrees of freedom, and $\hat{\cal H}_{\rm OM}$ contains the OM coupling between the photons and the vibrational mode.
The position operator $\hat{X}_{\ell}$ for the membranes, with respect to the equilibrium position, reads $\hat{X}_{\ell} = X_{\rm M} (\hat{b}_{\ell}^{\dag} + \hat{b}_{\ell})$ in terms of the lowering operators $\hat{b}_{\ell}$ for the mechanical modes, where $X_{\rm M} = \sqrt{1/(2 m \Omega_{\rm M})}$.
The Hamiltonian reads
\begin{equation}\label{eq:mechanicalhamil}
\hat{\cal H}_{\rm M} = \sum_{\ell=1}^{2} \Omega_{\rm M} \hat{b}_{\ell}^{\dag} \hat{b}_{\ell} + \sum_{\ell=1}^{2} F_{\ell}(t) \hat{X}_{\ell}~.
\end{equation}
The membranes can be subjected to periodic driving $F_{\ell}(t) = F(e^{+i \Omega t + i \varphi_{\ell}} + \mbox{c.c.})$ with frequency $\Omega$, force $F$, and phase $\varphi_{\ell}$.
For simplicity, we assume that only the phase of the driving is membrane-dependent.
The photonic part reads
\begin{eqnarray}\label{eq:photonhamil}
\hat{\cal H}_{\rm ph} & = & \omega \hat{a}_{1}^{\dag} \hat{a}_{1} + \omega \hat{a}_{2}^{\dag} \hat{a}_{2} + \omega' \hat{c}^{\dag} \hat{c} \nonumber \\
& & - J ( \hat{a}_{1}^{\dag} \hat{c} + \hat{c}^{\dag} \hat{a}_{1} +
\hat{a}_{2}^{\dag} \hat{c} + \hat{c}^{\dag} \hat{a}_{2})~,
\end{eqnarray}
where $\hat{c}$ and $\hat{a}_{\ell}$ are lowering operators for the photonic modes localized between the two membranes (with frequency $\omega'$), and between the membranes and the mirrors (with frequency $\omega$), respectively.
From the very rich spectrum of the system, here we have singled out two frequencies only, such that the difference $\delta\omega = \omega' - \omega$ is almost resonant with the membrane frequency $\Omega_{\rm M}$, with detuning $\Delta_{\rm M} = \Omega_{\rm M} - \delta\omega$.
We remark here that $\delta \omega$ is not related to the free spectral range of the empty cavity, and the equilibrium position of the membranes can be chosen to fulfill the resonance condition, exploiting the large number of avoided crossings between optical modes~\cite{TZJMGH2008,SYZJH2010}.
We assume here that a suitable level structure is found in the neighborhood of an avoided crossing, and that the rest of the spectrum can be neglected, in analogy to the two-level approximation commonly employed in atomic physics.

The frequency of the mode $\hat{a}_{\ell}$ depends on the position operator $\hat{X}_{\ell}$, while the frequency of $\hat{c}$ depends on the difference $\hat{X}_{1} - \hat{X}_{2}$.
Upon expanding such dependence to first order, one finds the linear dispersive coupling between the photonic modes and the membranes.
The resulting Hamiltonian reads
\begin{equation}\label{eq:optomechcoupling}
\hat{\cal H}_{\rm OM} = - g \hat{a}_{1}^{\dag} \hat{a}_{1} \hat{X}_{1} + g \hat{a}_{2}^{\dag} \hat{a}_{2} \hat{X}_{2} + g' \hat{c}^{\dag} \hat{c} ~ (\hat{X}_{1} - \hat{X}_{2})~.
\end{equation}
Finally, the interaction of the vibrational mode of the membranes with all the other mechanical modes of the structure is modeled by the Liouvillian
\begin{equation}\label{eq:liouvphonons}
{\cal L}[\rho(t)] = \sum_{\ell=1}^{2} \left \lbrace \frac{\gamma}{2} (n_{\rm th} + 1)
\Lambda[\hat{b}_{\ell}][\rho(t)] + \frac{\gamma}{2} n_{\rm th} \Lambda[\hat{b}_{\ell}^{\dag}][\rho(t)] \right \rbrace~.
\end{equation}
When the mechanical system is in equilibrium with its environment, $n_{\rm th}$ is the occupation of the phononic bath at the energy $\Omega_{\rm M}$, and $\gamma$ is the average damping.
However, $n_{\rm th}$ and $\gamma$ should be understood as effective parameters in the presence of laser cooling and, in particular, $\gamma$ can be increased and $n_{\rm th}$ reduced below unity~\cite{WNZK2007,MCCG2007,GVTGA2008}.

The effect of driving the membranes is to induce transitions between photonic states with different frequencies.
To show this, we perform a unitary transformation $\hat{U} \hat{\cal H} \hat{U}^{\dag}$ of the Hamiltonian, such that the photonic fields, to linear order in $J/\delta\omega$, transform as
\begin{eqnarray}\label{eq:trasfomodes}
\hat{a}_{1} \mapsto \hat{a}_{1} - (J / \delta\omega) \hat{c},\quad
\hat{a}_{2} \mapsto \hat{a}_{2} - (J / \delta\omega) \hat{c}, \nonumber \\
\hat{c} \mapsto -\hat{c} - (J / \delta\omega) \hat{a}_{1} - (J / \delta\omega) \hat{a}_{2}~.
\end{eqnarray}
(In the following, only the transformed operators will be used so, to avoid cumbersome notation, we do not introduce new symbols.)
The transformation has been chosen to diagonalize the hopping part of the Hamiltonian (\ref{eq:photonhamil}) that, in terms of the new operators, reads
\begin{equation}\label{eq:photonhamiltrasf}
\hat{\cal H}_{\rm ph} = \omega \hat{a}_{1}^{\dag} \hat{a}_{1} + \omega \hat{a}_{2}^{\dag} \hat{a}_{2} + \omega' \hat{c}^{\dag} \hat{c} + {\cal O}[(J / \delta\omega)^{2}]~.
\end{equation}
The quadratic part in $J/\delta\omega$, omitted from the Hamiltonian for the moment, contains hopping between the degenerate modes $\hat{a}_{\ell}$.
The OM coupling (\ref{eq:optomechcoupling}) reads
\begin{eqnarray}\label{eq:optomechhamiltrasf}
\hat{\cal H}_{\rm OM} & = & -g \hat{a}_{1}^{\dag} \hat{a}_{1} \hat{X}_{1} + g \hat{a}_{2}^{\dag} \hat{a}_{2} \hat{X}_{2}  + g' \hat{c}^{\dag} \hat{c} ~( \hat{X}_{1} - \hat{X}_{2} ) \nonumber \\
& & + (J / \delta\omega) \hat{c}^{\dag} [ g' \hat{a}_{2} \hat{X}_{1} - g' \hat{a}_{1} \hat{X}_{2} + (g + g') \hat{a}_{1} \hat{X}_{1} \nonumber \\
& & - (g + g') \hat{a}_{2} \hat{X}_{2} ] + \mbox{H.c.} + {\cal O}[(J / \delta\omega)^{2}]~.
\end{eqnarray}
In the following, we neglect the first line of (\ref{eq:optomechhamiltrasf}), which contains non-resonant terms that account only for a renormalization of the bare frequencies in Eq.~(\ref{eq:photonhamiltrasf}) (see also Ref.~\cite{GLPV2010}).
The remaining lines, on the contrary, represent a ``three-wave mixing'' effect, in which the energy of a phonon is absorbed (emitted) to scatter a photon to a state of higher (lower) energy.
In the present derivation, the phonon-mediated coupling between modes with different energy arises because of the mode mixing due to the hopping terms in the original Hamiltonian.
A similar approach has been considered in Ref.~\cite{HHM2010}.
A more rigorous derivation, starting from an {\it ad hoc} quantization of the photonic modes in the presence of a moving membrane, has been provided in Ref.~\cite{CL2011}.

It is now convenient to perform a displacement $\hat{b}_{\ell} \mapsto (X_{\ell}/X_{\rm M}) e^{-i \Omega t} + \hat{b}_{\ell}$ of the lowering operators of the mechanical mode, where $X_{\ell} = e^{-i \varphi_{\ell}} F X_{\rm M}^{2} / (\Omega - \Omega_{\rm M} + i \gamma / 2)$.
The displacements removes the driving term from the Hamiltonian~(\ref{eq:mechanicalhamil}), in the rotating-wave approximation.
In terms of the displaced fields, the Liouvillian (\ref{eq:liouvphonons}) is invariant in form, while the OM Hamiltonian~(\ref{eq:optomechhamiltrasf}) can be decomposed as $\hat{\cal H}_{\rm OM} = \hat{\cal H}_{\rm D} + \hat{\cal H}_{\rm vib}$, where
\begin{equation}\label{eq:hamildriving}
\hat{\cal H}_{\rm D} = {\cal E} e^{-i \Omega t} \hat{c}^{\dag} (\hat{a}_{2} - \hat{a}_{1}) + \mbox{H.c.}~,
\end{equation}
with the effective driving strength ${\cal E} = - g J X_{1} / \delta\omega$, and
\begin{eqnarray}
\hat{\cal H}_{\rm vib} & = & (J / \delta\omega) X_{\rm M} \left \lbrace  \hat{c} [g' \hat{a}_{2}^{\dag} + (g + g') \hat{a}_{1}^{\dag}] \hat{b}_{1}^{\dag} \right . \nonumber \\
& & \left . - \hat{c} [g' \hat{a}_{1}^{\dag} + (g + g') \hat{a}_{2}^{\dag}] \hat{b}_{2}^{\dag} + \mbox{H.c.} \right \rbrace~.
\end{eqnarray}
The first part describes photonic transitions induced by the membrane driving, while the second part contains the coupling between the photons and the quantized vibrational mode.
In Eq.~(\ref{eq:hamildriving}) we also made the specific choice $\varphi_{\ell} = 0$, which means that the forced vibration is actuated with the same phase on both membranes.
We see that, with this choice of phases, a state $|\Psi\rangle \propto (\hat{a}_{1}^{\dag} + \hat{a}_{2}^{\dag})^{N} |{\rm vac}\rangle$ [cfr. Eq.~(\ref{eq:becstate})], where all the photonic population is contained in the mode given by the symmetric combination of the modes $\hat{a}_{1}$ and $\hat{a}_{2}$, is unaffected by the Hamiltonian $\hat{\cal H}_{\rm D}$.
In this sense, the symmetric state is dark with respect to the mechanical driving, which excites only photons in the antisymmetric mode to the upper mode $\hat{c}$.
Finally, it is convenient to remove the rotating phase from the Hamiltonian (\ref{eq:hamildriving}) using a new frame for the photonic operators, such that the photonic Hamiltonian (\ref{eq:photonhamiltrasf}) reads $\hat{\cal H}_{\rm ph} = -\Delta \hat{c}^{\dag}\hat{c}$, with $\Delta = \Omega - \delta\omega$, and $\hat{\cal H}_{\rm D} = {\cal E} \hat{c}^{\dag} (\hat{a}_{2} - \hat{a}_{1}) + \mbox{H.c.}$.

Resonant ``three-wave-mixing'' processes in OM systems have been previously considered for the transduction of photons and phonons~\cite{CSHP2011} and have been proposed for enhancing nonlinear effects in strongly coupled OM systems~\cite{LSPM2012,SKHBLZR2012}.
Here we are interested in the regime where the mechanical modes, which are involved in this process, are strongly damped and therefore act as a reservoir for the photons.
Assuming that the coupling strength between the photonic and phononic modes is smaller than the decay width of the phonons, i.e.~$J|g|/\delta\omega,J|g'|/\delta\omega \ll \gamma$, we perform the Born-Markov approximation and trace out the phononic modes $\hat{b}_{\ell}$.
As a result the reduced density matrix $\rho_{\rm ph}(t) = {\rm Tr}_{\rm M}[\rho(t)]$ for the photons obeys the ME (\ref{eq:masterequation}) with the Hamiltonian $\hat{\cal H} = \hat{\cal H}_{\rm ph} + \hat{\cal H}_{\rm D}$ and the Liouvillian ${\cal L} = {\cal L}_{\rm eff}$, where
\begin{eqnarray}\label{eq:effliouv}
{\cal L}_{\rm eff}[\rho_{\rm ph}(t)] & = & \sum_{\ell=1}^{2}
\tilde{\Gamma} (n_{\rm th} + 1) \Lambda[\hat{K}_{\ell}] [\rho_{\rm ph}(t)]
\nonumber \\
& & + \sum_{\ell=1}^{2} \tilde{\Gamma} n_{\rm th} \Lambda[\hat{K}_{\ell}^{\dag}][\rho_{\rm ph}(t)]~,
\end{eqnarray}
with the jump operators
\begin{eqnarray}
\hat{K}_{1} & = & \hat{c} [\hat{a}_{2}^{\dag} + (1 + g / g') \hat{a}_{1}^{\dag}]~,\nonumber \\
\hat{K}_{2} & = & \hat{c} [\hat{a}_{1}^{\dag} + (1 + g / g') \hat{a}_{2}^{\dag}]~.
\end{eqnarray}
The effective damping rate in the Liouvillian~(\ref{eq:effliouv}) reads $\tilde{\Gamma} \simeq 2 (g' X_{\rm M} J / \delta\omega)^{2} / \gamma$.
We point out that our derivation holds at leading order in $J / \delta\omega$, i.e.~linear for the Hamiltonian and quadratic for the effective Liouvillian~(\ref{eq:effliouv}).
Higher-order corrections to the structure of the coupling Hamiltonian would yield corrections to orders higher than two to the Liouvillian, and hence can be neglected.
We also note that, in the limit $g' \gg g$, the two jump operators are equal and, after the adiabatic elimination of the mode $\hat{c}$ (see Sec.~\ref{sec:projectiontechnique}), they reproduce the form (\ref{eq:paradigmaticjump}) that was studied in the atomic context~\cite{DMKKBZ2008}.

Finally, coherent excitations of  the cavity modes by external laser fields can be described by
\begin{equation}
\hat{\cal H}_{\rm co} = \sum_{\ell=1}^{2} \varepsilon_{\ell}(t)\hat{a}_{\ell}^{\dag} + \mbox{H.c.}~,
\end{equation}
with $\varepsilon_{\ell}(t) = \varepsilon_{\ell} e^{-i \Omega_{\ell} t}$ in the rotating-wave approximation.
Losses of photons through the mirrors of the cavity are described by the Liouvillian
\begin{equation}
{\cal L}_{\rm loss}[\rho_{\rm ph}(t)] = \sum_{\ell=1}^{2} \kappa
\Lambda[\hat{a}_{\ell}][\rho_{\rm ph}(t)] + \kappa \Lambda[\hat{c}][\rho_{\rm ph}(t)]~,
\end{equation}
where we assume, for simplicity, that the decay width $\kappa$ is the same for all modes.

In the following, we omit the subscript in the photonic density matrix and redefine $\rho_{\rm ph}\rightarrow \rho$, which all together obeys the ME
\begin{equation}\label{eq:PhotonME}
\dot \rho= -i [ \hat{\cal H}_{\rm ph} + \hat{\cal H}_{\rm D} + \hat{\cal H}_{\rm co}, \rho]Ê +  {\cal L}_{\rm eff} \rho + {\cal L}_{\rm loss}\rho.
\end{equation}
It describes the open system dynamics of the two cavity modes $\hat a_1$ and $\hat a_2$, which are externally driven and coupled dissipatively via excitation and successive decay of the intermediate mode $\hat c$.

\subsection{Adiabatic elimination of the excited mode}
\label{sec:projectiontechnique}

The full model for the optical modes given in Eq.~(\ref{eq:PhotonME}) can be further simplified if the population in the excited mode $\hat{c}$ is negligibly small.
This is indeed the case for appropriate choices of parameters, as we point out in the following sections.
In this case, we may proceed to the adiabatic elimination of the mode $\hat{c}$ to obtain an effective ME for the modes $\hat{a}_{1}$ and $\hat{a}_{2}$ only.

We start out by introducing the projection super-operators~\cite{BreuerPetruccioneBook} ${\cal P}\rho \equiv {\rm Tr}_{c}[\rho] \otimes |0\rangle_{c} \langle 0|$ onto the vacuum state of mode $\hat{c}$ and its complement ${\cal Q} \equiv \mathbb{1} - {\cal P}$.
By neglecting the weak external driving and the cavity losses (which just lead to an overall decay of all modes) in the present derivation, the equations of motion (EOM) for the projections read
\begin{eqnarray}\label{eq:evolprojected}
\partial_{t} {\cal P}\rho & = & {\cal P}{\cal L}_{\rm D}{\cal Q} \rho + {\cal P}{\cal L}_{\rm eff}{\cal Q}\rho~, \nonumber \\
\partial_{t} {\cal Q}\rho & = & {\cal Q}({\cal L}_{\rm ph} + {\cal L}_{\rm eff} + {\cal L}_{\rm D}){\cal Q}\rho + {\cal Q}{\cal L}_{\rm D} {\cal P} \rho~,
\end{eqnarray}
where ${\cal L}_{\rm D}$ and ${\cal L}_{\rm ph}$ are the Liouvillians corresponding to $\hat{\cal H}_{\rm D}$ and $\hat{\cal H}_{\rm ph}$, respectively.
For simplicity we consider the limit $g' \gg g$ and $n_{\rm th} = 0$.
Note that here the ${\cal P}$ and ${\cal Q}$ subspaces are coupled both by the coherent linear driving in ${\cal L}_{\rm D}$ and the nonlinear decay in ${\cal L}_{\rm eff}$.
Our goal is to solve the EOM for ${\cal Q}\rho$ perturbatively in the driving strength ${\cal E}$, without assuming that the decay is small.
To this end, we decompose ${\cal Q}\rho(t) = Q_{1}(t) + Q_{2}(t)$ in a first- and second-order contribution, which are obtained by formally integrating Eq.~(\ref{eq:evolprojected}) and read
\begin{eqnarray}
Q_{1}(t) = \int_{-\infty}^{t} d\tau e^{({\cal L}_{\rm ph} + {\cal L}_{\rm eff})(t - \tau)}{\cal Q} {\cal L}_{\rm D} {\cal P}\rho(t), \nonumber \\
Q_{2}(t) = \int_{-\infty}^{t} d\tau e^{({\cal L}_{\rm ph} + {\cal L}_{\rm eff})(t - \tau)}{\cal L}_{\rm D} Q_{1}(\tau)~.
\end{eqnarray}
These expressions are inserted into the equation of motion for ${\cal P}\rho$ and the exponentials are explicitly evaluated using the decomposition
\begin{equation}
{\cal P}\rho = \sum_{n_{\rm s},n_{\rm s}'} \rho^{(a)}_{n_{\rm s},n_{\rm s}'} \otimes | n_{\rm s} \rangle \langle n_{\rm s}'|~,
\end{equation}
where the indices s and a refer to the symmetric $\hat a_{\rm s}= (\hat{a}_{1} + \hat{a}_{2})/\sqrt{2}$ and anti-symmetric $\hat a_{\rm a}=(\hat{a}_{1} - \hat{a}_{2})/\sqrt{2}$ mode, respectively.
As a result we obtain an effective ME for the reduced density operator of the $\hat a$ modes,  $\rho_{\rm a}= {\rm Tr} \{ {\cal P}\rho\}$, which can be written in the form
\begin{equation}
\dot \rho_{\rm a} = i  \left[\frac{2\mathcal{E}^2 \Delta }{\Delta^2 + \hat \Gamma^2} \hat a_{\rm a}^\dag \hat a_{\rm a}, \rho_a\right] + 8 \, \mathcal{E}^2 \tilde{\Gamma} \Lambda[\hat K] \rho_{\rm a} + \mathcal{L}_{\rm loss}\rho_{\rm a},
\end{equation} 
where $\hat \Gamma= 4 \tilde{\Gamma}(\hat a_{\rm s}^\dag \hat a_{\rm s} +1)$ and we have introduced the jump operator  
\begin{equation}
\hat K= \hat a_{\rm s}^\dag \left(\frac{ 1}{i\Delta + 4 \tilde{\Gamma} (\hat a_{\rm s}^\dag \hat a_{\rm s}+1)} \right)\hat a_{\rm a}.
\end{equation}
In the limit $\Delta \gg \tilde{\Gamma} \langle \hat a_{\rm s}^\dag \hat a_{\rm s}\rangle$, the Hamiltonian reduces to a shift of the antisymmetric mode, while the jump operator recovers the form (\ref{eq:paradigmaticjump}), i.e.~$\hat K \propto \hat{a}_{\rm s}^\dag \hat{a}_{\rm a}$.
For larger $\tilde \Gamma$, additional higher order nonlinearities arise, which, for example, can lead to pumping-induced anti-bunching effects, as described in Ref.~\cite{MTDR2012}.
 
In currently realized OM devices with membranes inside a Fabry-Perot cavity~\cite{TZJMGH2008,SYZJH2010} the optical free spectral range is in the ${\rm GHz}$ range and flexural mechanical modes have frequencies $\Omega_{\rm M} \simeq {\rm MHz}$. 
At the same time, very low cavity decay rates $\kappa \simeq 50\, {\rm kHz}\ll \Omega_{\rm M}$ can be achieved, such that the separation of frequency scales assumed in our derivation can be realized.
For typical mechanical quality factors of $Q \simeq 10^6$, the bare mechanical decay rate $\gamma_0$ can be as low as $1~{\rm Hz}$, but it can be tuned up to $\gamma \simeq \Omega_{\rm M}$ by applying additional laser cooling~\cite{MCCG2007,WNZK2007} or feedback damping~\cite{GVTGA2008,KB2006}. 
This would also reduce the effective temperature to mean occupations $n_{\rm th}\simeq \mathcal{O}(1)$.
Note that although in these systems the OM coupling is rather low, $g \simeq 1-10~{\rm Hz}$ and the resulting dissipation $\tilde \Gamma \simeq g^2/\gamma \ll \kappa$, the nonlinear effects we are interested in will depend on the ratio $\tilde \Gamma \langle \hat a_{\rm s}^\dag \hat a_{\rm s}\rangle/\kappa$, which can exceed unity for larger photon numbers. 
Furthermore, in several solid state realizations of OM devices, such as microtoroidal cavities or OM crystals [see Fig.~\ref{fig:lambdasystem} (c) and (d)],  OM coupling constants $g$ of few kHz up to $\simeq 1~{\rm MHz}$~\cite{VDWSK2012,DBSLDLF2010,CMSHKGAP2011} have been demonstrated, and even the single-photon strong-coupling regime $g > \kappa$ is within reach of optimized devices.
In such devices, the regime $\tilde \Gamma \simeq \kappa$ is in principle accessible,  while at the same time the high mechanical frequencies $\Omega_{\rm M} \simeq {\rm GHz}$~\cite{DBSLDLF2010,CMSHKGAP2011} lead to equilibrium occupation numbers $n_{\rm th} < 1$ at cryogenic temperatures.
This parameter regime will be considered in most of our numerical simulations below.

\subsection{Numerical results}

We study now in detail the building block of the many-body driven-dissipative system, i.e.~the effective $\Lambda$ system for photons introduced in Sec.~\ref{ssec:themodel}, according to Eq.~(\ref{eq:PhotonME}).
First we consider the case in which the lower mode $\hat{a}_{1}$ is coherently driven.
The coherent pumping is resonant with the single mode, which means that it is equally detuned from both the symmetric $(\hat{a}_{1} + \hat{a}_{2})/\sqrt{2}$ and the antisymmetric $(\hat{a}_{1} - \hat{a}_{2})/\sqrt{2}$ modes, which are split by $2 \tilde{J}$.
The splitting arises from the residual coupling in Eq.~(\ref{eq:photonhamiltrasf}), which we write $\hat{\cal H}_{\rm hopp} = -\tilde{J} (\hat{a}^{\dag}_{1} \hat{a}_{2} + \mbox{H.c.})$.

The EOM are solved using a standard fourth-order Runge-Kutta method, truncating the Fock space of each mode to 4--6 states.
The coherent pumping couples equally to the symmetric and antisymmetric mode, however, because of the OM driving and dissipation, the steady-state population in the symmetric mode is substantially larger than the population in the antisymmetric mode.
During the time-evolution, the population in the upper mode $\hat{c}$ can be made negligibly small, employing a  large detuning $\Delta$ of the OM driving.

\begin{figure}
\includegraphics[width=\linewidth]{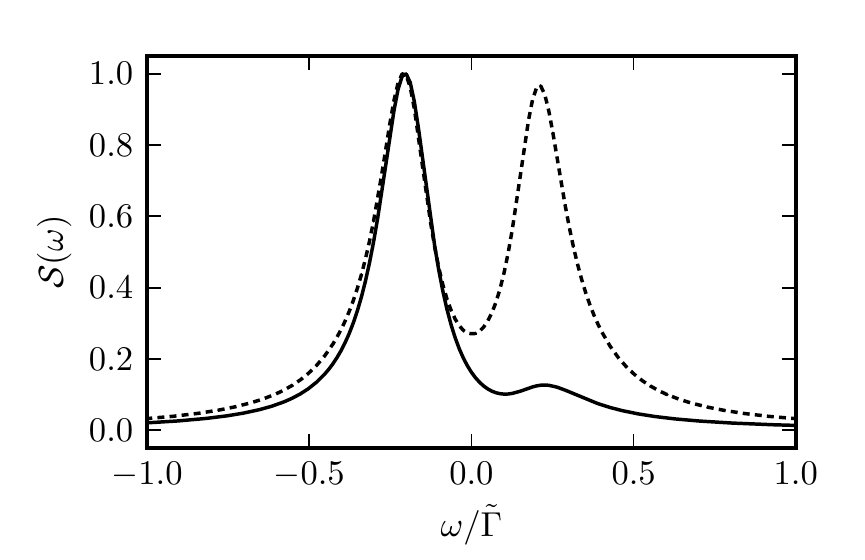}
\caption{\label{fig:twopeaks}
Spectrum of the light radiated from the $\hat{a}_{2}$ mode.
The $\hat{a}_{1}$ mode is coherently pumped with strength $\varepsilon_{1} = 0.1 \tilde{\Gamma}$, all the modes decay with amplitude $\kappa = 0.1 \tilde{\Gamma}$.
The OM driving is $ {\cal E} = 0.5 \tilde{\Gamma} $ (solid line) $ 0.01 \tilde{\Gamma} $ (dashed line), with $\tilde{J} = 0.2 \tilde{\Gamma}$.
The system is first evolved to the steady state at $t_{0} \tilde{\Gamma} = 50.0$ and subsequently to $t \tilde{\Gamma} = 500.0$ to compute the two-times correlation functions with sufficient spectral resolution.  }
\end{figure}
In Fig.~\ref{fig:twopeaks} we present a clear signature of the population imbalance obtained by means of the OM driving.
More precisely, we compute first the connected two-time Green's function
\begin{equation}
{\cal G}_{2}^{\rm (c)}(t,t_{0}) = \langle [\hat{a}_{2}^{\dag}(t) - \langle \hat{a}_{2}^{\dag}(t) \rangle][\hat{a}_{2}(t_{0}) - \langle \hat{a}_{2}(t_{0}) \rangle] \rangle
\end{equation}
in the mode $\hat{a}_{2}$, where the average is computed on the steady state at $t = t_{0}$.
The absolute value of the Fourier transform with respect to the time $t$ is then normalized to unity to give the cavity output spectrum ${\cal S}(\omega)$.
We see that, in the absence of OM driving, the spectrum is composed of two equal peaks which represent the symmetric and antisymmetric mode (at lower and higher frequency, respectively).
Switching the OM coupling on, the antisymmetric peak is strongly suppressed, due to the efficient scattering of photons into the symmetric mode.
The suppression of the peak corresponding to the antisymmetric mode is an unambiguous consequence of the phase-locking between the two sites and clearly distinguishes our setup from a more conventional situation where two independent cavities are pumped coherently, without being phase-locked to each other.
The effect of switching on the OM coupling is studied in more detail in App.~\ref{app:separability}, where the entanglement between the symmetric mode and the rest of the system is discussed as well.

\begin{figure}
\includegraphics[width=\linewidth]{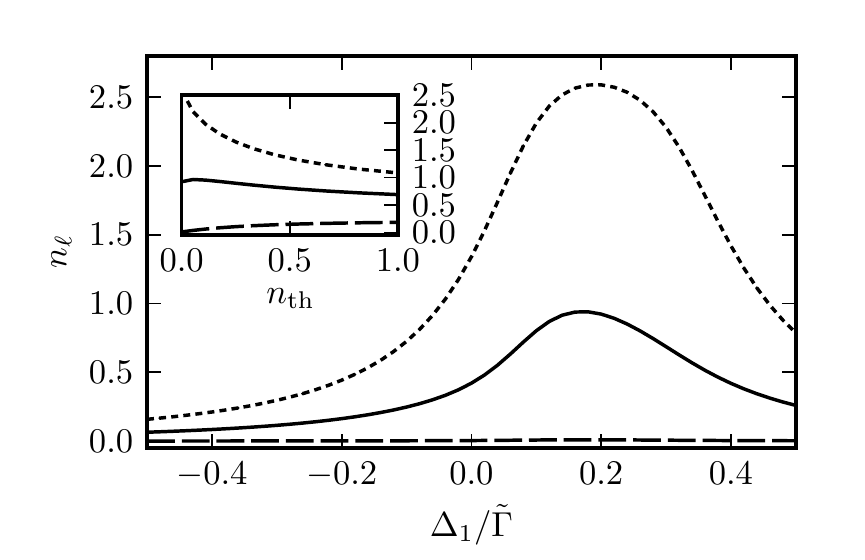}
\caption{\label{fig:detuneddriving}
Main panel: steady-state populations $n_{\ell}$ for the symmetric (solid line), antisymmetric (short dashed), and excited (long dashed) mode, in the presence of coherent pumping of the antisymmetric mode with detuning $\Delta_{1}$.
Inset: peak value of $n_{\ell}$ at detuning $\Delta_{1} = \tilde{J} = 2.0 \tilde{\Gamma}$ with finite thermal occupation $n_{\rm th}$ of the phononic reservoir.
The strength of the coherent pumping is $\varepsilon_{\ell} = 2.0 \kappa$, with ${\cal E} = 2.0 \tilde{\Gamma}$ and $\tilde{\Gamma} = 10.0 \kappa$. }
\end{figure}
While in Fig.~\ref{fig:twopeaks} we show the noise spectrum on one side of the system while both symmetric and antisymmetric modes are pumped coherently, in Fig.~\ref{fig:detuneddriving} we illustrate the effect of pumping the antisymmetric mode only, with variable detuning.
We use opposite Rabi frequencies in the two modes $\ell = 1$ and $\ell = 2$, so that the coupling to the symmetric mode vanishes.
The population in the steady state shows broad resonances centered around the frequency $\tilde{J}$ of the antisymmetric mode.
While the broadening of the noise spectrum in Fig.~\ref{fig:twopeaks}, at zero temperature, is determined by the photonic decay in each mode, the broadening in Fig.~\ref{fig:detuneddriving} is given by the larger scale $\tilde{\Gamma}$.
It is interesting to note that this procedure shows a resonance in the symmetric mode which is centered around the energy of the antisymmetric mode as well.
This is due to the fact that the excess energy is taken away by the phonons that are emitted following the (virtual) excitation of the mode $\hat{c}$.
The latter mode, as is apparent from the figure, is always negligibly populated.
Finally, the effect of the finite temperature of the phononic reservoir is investigated in the inset of Fig.~\ref{fig:detuneddriving}.
We see that the driven-dissipative mechanism is robust in the presence of low thermal occupations $n_{\rm th}$, as the magnitude of the peak corresponding to the symmetric state remains substantial.

\section{Membranes array}
\label{sec:membranesarray}

\begin{figure}
\includegraphics[width=\linewidth]{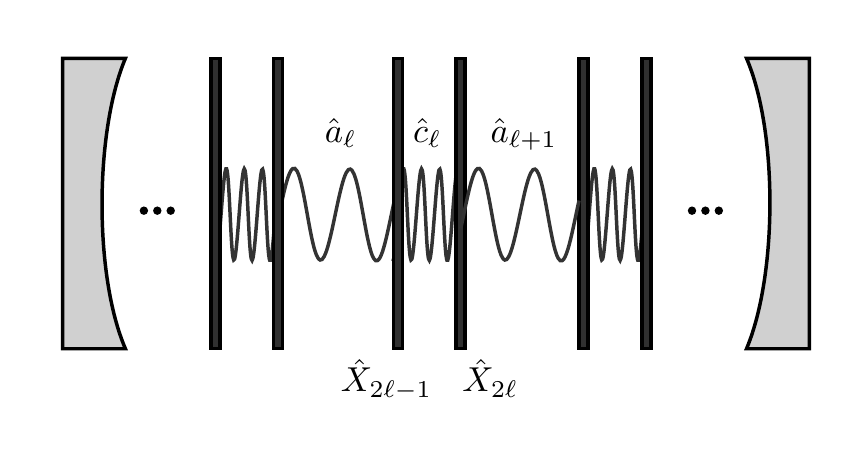}
\caption{\label{fig:setup}
Schematic representation of the system under consideration.
An array of membranes (thick vertical lines), grouped into doublets at positions $X_{2\ell - 1}$ and $X_{2\ell}$, within an optical cavity, define two sets of quantized localized light modes $\hat{a}_{\ell}$ and $\hat{c}_{\ell}$.
In Sec.~\ref{sec:twomembranes} we discuss the case in which only two membranes are present. }
\end{figure}

In this section we discuss the extension of the $\Lambda$-system for photons of Sec.~\ref{sec:twomembranes} to the case of an array of $2L$ membranes.
The model for the membranes array~\cite{BM2008,XGD2012} is a straightforward generalization of the previous case, and only the final results are reported.
The time-evolution of the extended system, however, is much more complex, because the photonic field is inhomogeneous.
The dissipative mechanism must single out the symmetric mode in the lattice [with zero quasimomentum $k = 0$, cfr.~Eq.~(\ref{eq:becstate})] among the many modes of the Brillouin zone, while only the central (symmetric) and edge (antisymmetric) Brillouin-zone modes are present in the two-membrane system.
We show numerically that the driven-dissipative dynamics achieves phase-locking of the photonic field throughout the array, i.e.~the photonic field converges towards the $k = 0$ state.

The membranes are disposed in a regular array of pairs at positions $X_{2\ell-1}$ and $X_{2\ell}$ (see Fig.~\ref{fig:setup}) and define two sets of localized  photonic modes ($\hat{a}_{\ell}$ and $\hat{c}_{\ell}$) with different frequencies ($\omega$ and $\omega'$, respectively).
Because of photon transmission through the membranes, the spectrum splits into two bands centered at the two frequencies $\omega$ and $\omega'$.
The splitting depends on the hopping amplitude $J$ between the modes, and we assume that it can be neglected with respect to the difference $\delta \omega = \omega' - \omega$.
The transformation (\ref{eq:trasfomodes}) now reads in general
\begin{equation}
\hat{a}_{\ell} \mapsto w \hat{c}_{\ell-1} + r \hat{a}_{\ell} + w
\hat{c}_{\ell},\quad
\hat{c}_{\ell} \mapsto w' \hat{a}_{\ell} + r' \hat{c}_{\ell} + w' \hat{a}_{\ell+1}~,
\end{equation}
and the following results hold to the first nonzero order in $w$ and $w'$.
A specific example of the transformation is given in App.~\ref{app:newmodes}.
The photonic part of the Hamiltonian, in the rotating frame, reads $\hat{\cal H}_{\rm ph} = - \sum_{\ell=1}^{L-1} \Delta \hat{c}_{\ell}^{\dag} \hat{c}_{\ell}$, while the driving part (\ref{eq:hamildriving}) becomes
\begin{equation}
\hat{\cal H}_{\rm D} = \sum_{\ell=1}^{L-1} \left [ (-)^{\ell} {\cal
    E}\hat{c}_{\ell}^{\dag}(\hat{a}_{\ell} - \hat{a}_{\ell+1}) + \mbox{H.c.} \right ]~.
\end{equation}
with ${\cal E} = 2 g X w^{\ast} r$.
Here we assume that each pair of membranes vibrates rigidly together, while neighboring pairs have opposite velocities.
More precisely, the phase of the driving is $\varphi_{2\ell-1} = \varphi_{2\ell} = 0$ if $\ell$ is odd, and $\varphi_{2\ell-1} = \varphi_{2\ell} = \pi$ if $\ell$ is even.
In the effective Liouvillian (\ref{eq:effliouv}), the summations extend to $\ell=L$ with the jump operators
\begin{eqnarray}\label{eq:jumpoperatorslattice}
\hat{K}_{2\ell-1} & = & \chi \hat{a}_{\ell}^{\dag} ( \hat{c}_{\ell-1} +
\hat{c}_{\ell}) + (\hat{a}^{\dag}_{\ell} +  \hat{a}_{\ell+1}^{\dag})
\hat{c}_{\ell}, \nonumber \\
\hat{K}_{2\ell} & = & \chi \hat{a}^{\dag}_{\ell+1} ( \hat{c}_{\ell}
+ \hat{c}_{\ell+1}) + (\hat{a}_{\ell}^{\dag} +
\hat{a}_{\ell+1}^{\dag}) \hat{c}_{\ell}~, \nonumber \\
\end{eqnarray}
with $\chi = g w^{\ast}r / (g' w' r'^{\ast})$ and the effective damping $\tilde{\Gamma} \simeq 2 |g' X_{\rm M} w' r' |^{2} / \gamma$.

\begin{figure}
\includegraphics[width=\linewidth]{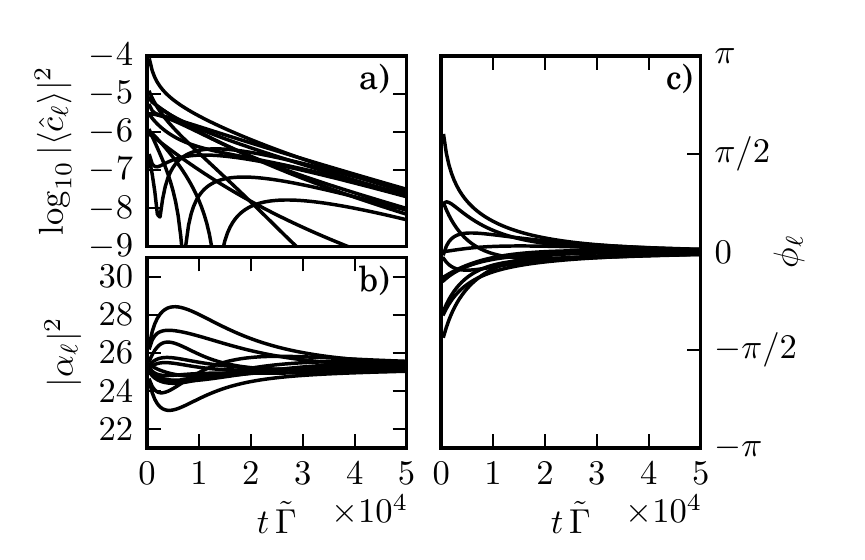}
\caption{\label{fig:semiclassical}
Time-evolution of the photonic population in the excited [panel (a)] and lower [panel (b)] levels, and of the relative phase [panel (c)] between the semiclassical field on neighboring ground sites, computed according to the average-number-conserving semiclassical EOM.
$L = 12$ sites (corresponding to $2 L$ membranes) are used with periodic boundary conditions.
The OM driving is $0.2 \tilde{\Gamma}$ with detuning $-10.0 \tilde{\Gamma}$. }
\end{figure}

We first investigate the time-evolution of an inhomogeneous initial state, in the absence of pumping and dissipation, in the semiclassical approximation.
To derive the semiclassical EOM, we first compute the EOM for the averages $\alpha_{\ell} = \langle \hat{a}_{\ell} \rangle$ and $\langle \hat{c}_{\ell} \rangle$ and we substitute each field operator with the corresponding $c$-number.
The solution of the EOM in Fig.~\ref{fig:semiclassical} shows that the system is driven towards a state where the photonic occupation in the $\hat{a}_{\ell}$ modes is homogeneous, while the occupation in the $\hat{c}_{\ell}$ modes decreases exponentially.
The relative phase $\phi_{\ell} = {\rm Arg}[\alpha_{\ell}\alpha_{\ell+1}^{\ast}]$ between neighboring $\hat{a}_{\ell}$ modes vanishes, demonstrating the onset of phase-locking in the array and the generation of the state with quasimomentum $k = 0$.

Given the exponentially suppressed occupation of the modes $\hat{c}_{\ell}$, visible from the solution of the semiclassical EOM and from the exact diagonalization discussed in Sec.~\ref{sec:twomembranes}, we may proceed to the adiabatic elimination of such modes.
This allows us in Sec.~\ref{sec:phasetransition} to focus on the many-body quantum dynamics of the modes $\hat{a}_{\ell}$ only.
The EOM for the operator $\hat{c}_{\ell}$ close to the the steady state is
\begin{equation}
\partial_{t} \hat{c}_{\ell}(t) \simeq i \Delta \hat{c}_{\ell}(t) -i (-)^{\ell} {\cal E} [\hat{a}_{\ell}(t) - \hat{a}_{\ell+1}(t)]~.
\end{equation}
Here we neglect the contribution of the dissipation $\tilde{\Gamma}$ with respect to the detuning $\Delta$, as demonstrated in Sec.~\ref{sec:projectiontechnique} in the case of two membranes.
We may hence substitute the steady-state form 
\begin{equation}
\hat{c}_{\ell}(t) \simeq (-)^{\ell} \frac{{\cal E}}{\Delta}[\hat{a}_{\ell}(t) -
  \hat{a}_{\ell+1}(t)]
\end{equation}
into the jump operators (\ref{eq:jumpoperatorslattice}).
The adiabatic approximation correctly captures that the average $\langle \hat{c}_{\ell}(t) \rangle$ vanishes when $\alpha_{\ell} = \alpha_{\ell+1}$, i.e.~on the steady state, as shown by the solution of the complete semiclassical equations in Fig.~\ref{fig:semiclassical}.
The effective jump operators for the modes $\hat{a}_{\ell}$ read
\begin{eqnarray}\label{eq:reducedjump}
\hat{K}_{2\ell-1} & = & (-)^{\ell}\frac{{\cal E}}{\Delta} \left [
(2 \chi + 1) \hat{a}_{\ell}^{\dag} \hat{a}_{\ell}
- \hat{a}_{\ell+1}^{\dag} \hat{a}_{\ell+1} \right . \nonumber \\
& & \left . -(1 + \chi) \hat{a}_{\ell}^{\dag} \hat{a}_{\ell+1}
+ \hat{a}_{\ell+1}^{\dag} \hat{a}_{\ell}
-\chi \hat{a}_{\ell}^{\dag} \hat{a}_{\ell-1} \right ], \nonumber \\
\hat{K}_{2\ell} & = & (-)^{\ell}\frac{{\cal E}}{\Delta} \left [
 \hat{a}_{\ell}^{\dag} \hat{a}_{\ell} -(2\chi
+ 1) \hat{a}_{\ell+1}^{\dag} \hat{a}_{\ell+1} \right . \nonumber \\
& & \left . + (\chi + 1) \hat{a}_{\ell+1}^{\dag} \hat{a}_{\ell}
- \hat{a}_{\ell}^{\dag} \hat{a}_{\ell+1}
+\chi \hat{a}_{\ell+1}^{\dag} \hat{a}_{\ell+2} \right ]~. \nonumber \\
\end{eqnarray}
We notice that these jump operators are considerably more involved than (\ref{eq:paradigmaticjump}) as they couple up to three neighboring sites each.
Each jump operator $\hat{K}_{\ell}$ describes the dissipation arising from the phonons in the $\ell$th membrane.
Dissipation processes due to different membranes are of course independent, and are then described by different Lindblad terms (\ref{eq:genericlindblad}).
The existence of a dark state when $n_{\rm th} = 0$ (i.e.~when the temperature of the phononic reservoir vanishes) can be easily seen in momentum space, where the jump operators read (apart from an irrelevant overall sign)
\begin{eqnarray}
\hat{K}_{2\ell-1} & = & \frac{1}{L} \sum_{q,k} \hat{a}^{\dag}_{q} \hat{a}_{k}
(-1)^{\ell} \frac{{\cal E}}{\Delta} e^{i k (\ell - 1)} e^{- i q (\ell + 1)}  \nonumber \\
& & \times [e^{i (k + q)} (1 + \chi) + e^{i k} - e^{i q} \chi]
[e^{i k} - 1]~, \nonumber \\
\hat{K}_{2\ell} & = & \frac{1}{L} \sum_{q,k} \hat{a}^{\dag}_{q} \hat{a}_{k}
(-1)^{\ell} \frac{{\cal E}}{\Delta} e^{i k \ell} e^{- i q (\ell + 1)} \nonumber \\
& & \times [1 + \chi - e^{i k} \chi + e^{i q} \chi]
[e^{i k} - 1]~,
\end{eqnarray}
with $\hat{a}_{\ell} = \sum_{q} e^{+i q \ell} \hat{a}_{q} / \sqrt{L}$.
Indeed, we see that the state with momentum $k=0$, which is the analogous in the lattice of the symmetrized mode $(\hat{a}_{1} + \hat{a}_{2}) / \sqrt{2}$ considered in Sec.~\ref{sec:twomembranes}, belongs to the kernel of the jump operator, and hence its population is never depleted by the driven-dissipative dynamics.

\section{Dynamical phase transition}
\label{sec:phasetransition}

In this section we consider the dynamics of the membrane array, introduced above, under continuous incoherent pumping and photonic losses.
In the steady state, the light intensity is governed by the rate of pumping and losses only.
However, the light coherence between different sites depends dramatically on the strength of the OM driving and vanishes below a critical point.
We argue that this effect can indeed be interpreted as a dynamical phase transition, which can be used to probe the effectiveness of the driven-dissipative mechanism in a strongly nonequilibrium setup.

We first discuss the Gutzwiller approximation, which is used to tackle the highly nontrivial problem of the time-evolution of the many-body system.
Then, we present the numerical solution to the EOM and, finally, a compact analytical method is presented, which yields a global picture for the the steady-state light coherence properties.

\subsection{Mean-field approximation}
\label{ssec:gutzwiller}

To investigate the quantum dynamics induced by the jump operators (\ref{eq:reducedjump}) beyond the semiclassical approximation used in Sec.~\ref{sec:membranesarray}, we resort to a mean-field approximation where the density matrix $\rho$ is factorized in real space into a product $\rho \simeq \bigotimes_{\ell} \rho_{\ell}$ of reduced local density matrices $\rho_{\ell}={\rm Tr}_{\neq \ell} \rho$ on the $\ell$th site~\cite{DTMFZ2010,TDZ2011}.
We remark that this approximation goes beyond the semiclassical approximation, as it conserves the full quantum structure of the local Fock space, so that local correlation functions are \emph{not} factorized.
This method has been successfully employed in Refs.~\cite{DTMFZ2010,TDZ2011} to study the time-evolution of number-conserving driven-dissipative systems.
The hypotheses that support the usage of the mean-field decoupling are valid in the present case as well, because the addition of local pumping and dissipation to the theory does not crucially rely on the representation of the nonlocal correlation functions, which are severely approximated in this mean-field scheme.
The equation of motion for the reduced density matrix $\rho_{\ell}(t)$ follows from
\begin{equation}
\partial_{t}\rho_{\ell}(t) = {\rm Tr}_{\neq \ell}[{\cal L}[\rho(t)]] \equiv {\cal L}_{\ell}[\rho_{\ell}(t)]~,
\end{equation}
using the expression detailed in App.~\ref{app:coefficients}.
In the following notation, the index of the site can be omitted from the operators, because the problem has been reduced to a collection of (nonlinear) local problems.
We also presently reduce to the homogeneous case, i.e.~we assume that the average $\langle \dots \rangle_{\ell}$ on the $\ell$th site is independent of $\ell$.
The mean-field approximation applied to the Liouvillian with jump operators (\ref{eq:reducedjump}) gives
\begin{eqnarray}\label{eq:mfliouv}
{\cal L}_{\rm eff}[\rho(t)] & = & \sum_{i,j=1}^{4} \left [ \Gamma (n_{\rm th}  + 1) L_{ij}^{(1)} + \Gamma n_{\rm th} L_{ij}^{(2)} \right ] \nonumber \\
& & \times [ 2 \hat{A}_{i}^{\dag} \rho(t) \hat{A}_{j} - \{ \dots , \rho(t) \} ]~,
\end{eqnarray}
with $\Gamma = \tilde{\Gamma} {\cal E}^{2} / \Delta^{2}$ and $\hat{A} = (1,\hat{a},\hat{a}^{\dag},\hat{a}^{\dag}\hat{a})$.
In the following, we focus on the zero-temperature limit $n_{\rm th} = 0$.
The $4\times 4$ matrices $L^{(1,2)}$ are reported in App.~\ref{app:coefficients}.
They depend on the expectation value of several (at most cubic) products of creation and annihilation operators and are functions of the density matrix $\rho$.
The equation of motion for $\rho$ is hence nonlinear in the mean-field approximation.
For $n_{\rm th} = 0$, the existence of the dark state with vanishing quasimomentum $k = 0$ is reproduced due to the fact that the Liouvillian vanishes when applied to the pure coherent state.

The photonic modes $\hat{a}$ are subjected to local losses at rate $\kappa$, which reduce the total particle number.
To prevent the photonic population to vanish in the steady state, photons are injected into the system at rate $\Upsilon$ by means of an incoherent pumping scheme.
The Liouvillian operators for such processes read
\begin{equation}\label{eq:incoherentpumping}
{\cal L}_{\rm pump}[\rho(t)] = \Upsilon \Lambda[\hat{a}^{\dag}][\rho(t)],\quad
{\cal L}_{\rm loss}[\rho(t)] = \kappa \Lambda[\hat{a}][\rho(t)]~.
\end{equation}
In the following, we consider the full dynamics in the array in the presence of the number-conserving Liouvillian (\ref{eq:mfliouv}) and of the pump and loss terms (\ref{eq:incoherentpumping}).
The latter terms induce a continuous flow of photons through the system, which has a substantial impact on the properties of the steady state.
Indeed, the effect of the Liouvillian~(\ref{eq:mfliouv}) goes towards the creation of a state featuring long-range phase locking in the array, while the pump and loss terms induce local dephasing.
The system exhibits then a competition between the two contributions to the non-unitary dynamics~\cite{VWC2009}.
Such competition results in a dynamical phase transition, as it is shown in the next section.

\subsection{Onset of coherence in the presence of incoherent pumping}
\label{ssec:phasetransition}

\begin{figure}
\includegraphics[width=\linewidth]{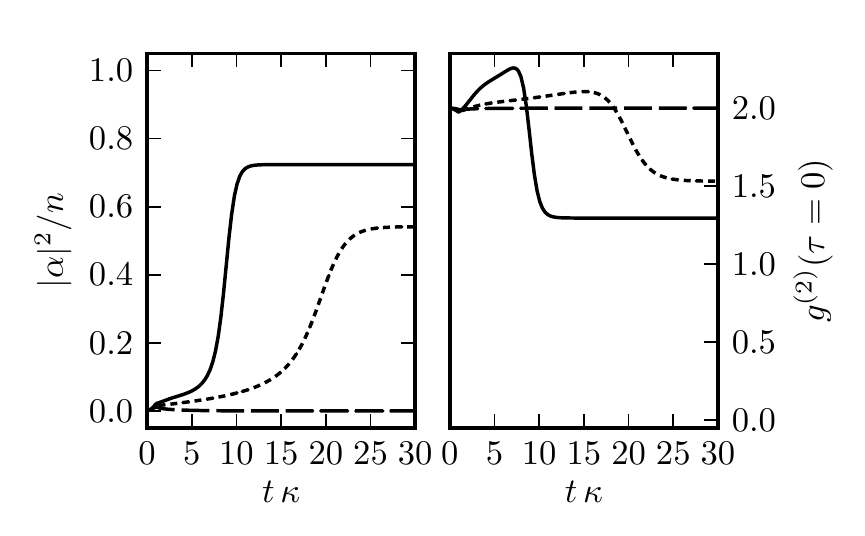}
\caption{\label{fig:evolarray}
Time-evolution of the coherent fraction $|\alpha|^{2}/n$ (left panel) and second order correlation function $g^{(2)}$ at zero delay (right panel), with incoherent pumping $\Upsilon = 0.5 \kappa$, for three different values of the rescaled driving strength $\Gamma' \simeq 0.5 \kappa$ (long dashed), $\Gamma' \simeq 1.5 \kappa$ (short dashed), and $\Gamma' = 2.5 \kappa$ (solid line), at zero temperature $n_{\rm th} = 0.0$.
The system is perturbed for $t \kappa \lesssim 1.0$ by a weak coherent pumping field.  }
\end{figure}

In the absence of the dissipative OM coupling $\Gamma = 0$, the combined effect of incoherent pumping and photon losses is to create a thermal state~\cite{GardinerZollerBook} with steady-state density $n_{\infty} = \Upsilon / (\kappa - \Upsilon)$, where we require $\Upsilon < \kappa$ for the system to be stable.
The density matrix of the thermal state is diagonal and the average $\alpha = \langle \hat{a} \rangle$ vanishes, indicating the absence of coherence.
The ``coherent fraction'' $|\alpha|^{2}/n$ achieves the maximal value $1$ in the coherent state, which is assumed in the semiclassical approximation.
In the Gutzwiller mean-field approximation for the density matrix, both these limiting cases are contained: the Fock space onsite density matrix can describe both a thermal state with matrix elements $\bar{\rho}_{n,n'} = p_n\delta_{n,n'}$, $p_n = \bar{n}^{n}/(\bar{n} + 1)^{n+1}$ as well as a coherent state $\rho = |\Psi\rangle_\text{coh}\langle \Psi|$,  $|\Psi\rangle_\text{coh} = \sum_{n=0}^{\infty}e^{-|\alpha|^{2} / 2}(\alpha^{n} / \sqrt{n !})|n\rangle$ indicating the presence of off-diagonal order. The coherent fraction measures the long-range coherence in the array, because the correlation functions factor as $\langle \hat{a}_{\ell'}^{\dag} \hat{a}_{\ell} \rangle \simeq \langle \hat{a}_{\ell'} \rangle^{\ast} \langle \hat{a}_{\ell} \rangle$ on the coherent state.
In this sense, the ratio $|\alpha|^{2}/n$ corresponds to the condensate fraction in the mean-field theory of weakly interacting Bose-Einstein condensates (see e.g.~Ref.~\cite{Zwerger2003}).

\begin{figure}
\includegraphics[width=\linewidth]{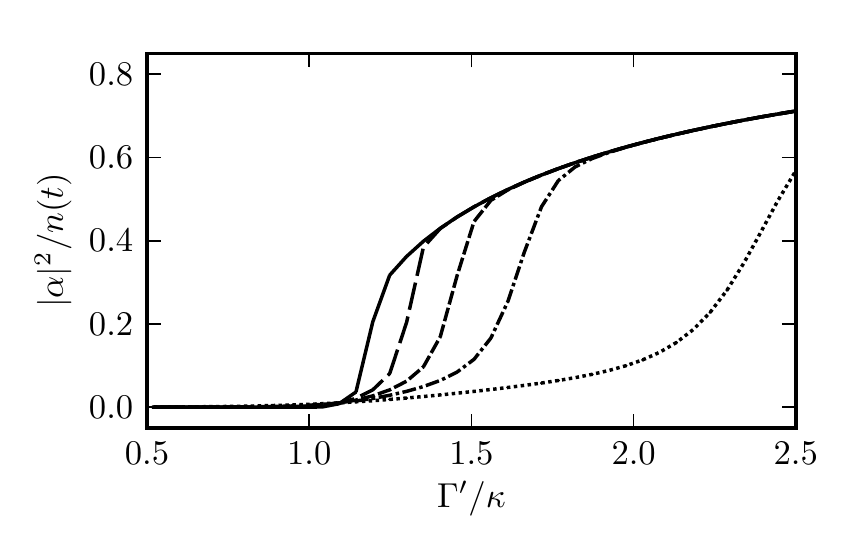}
\caption{\label{fig:phasetransition}
Dependence of the coherent fraction $|\alpha|^{2}/n$ on the rescaled driving strength $\Gamma'$, at zero temperature $n_{\rm th} = 0.0$, for $\Upsilon = 0.5 \kappa$.
The system is initialized in the vacuum state, perturbed for $t \kappa \lesssim 1.0$ by a weak coherent pumping field.
Stroboscopic plots of the profile at increasing times (dots, dash-dots, short dashes, long dashes, solid line) are shown to converge towards a steady-state profile.  }
\end{figure}

The main result of the present section is that a critical value $\Gamma_{\rm crit}' \simeq  \kappa$ exists, above which the condensate fraction is finite.
(The rescaled driving strength $\Gamma'$ is introduced in App.~\ref{app:ansatz}.)
The phase-locking effect engineered into the Liouvillian (\ref{eq:mfliouv}) overcomes the phase randomization induced by the photon losses and the pumping (\ref{eq:incoherentpumping}), generating a state with long-range order in the array.
This result is first illustrated in Fig.~\ref{fig:evolarray}, where we show the time-evolution of the coherent fraction for several values of the coupling $\Gamma'$.
We also show the second-order correlation function at zero delay $g^{(2)}(\tau=0) = \langle \hat{a}^{\dag} \hat{a}^{\dag} \hat{a} \hat{a} \rangle / n^{2}$ for the same parameters.
As the coherent fraction increases, $g^{(2)}$ decreases from $2$ towards $1$, signaling a transition from a thermal to a coherent state.
The steady-state value of the coherent fraction as a function of the driving strength $\Gamma'$ is finally shown in Fig.~\ref{fig:phasetransition}, which can be seen as a nonequilibrium phase diagram in one variable.
The incoherent and coherent phases are defined by the ``order parameter'' $\alpha$ being zero and nonzero, respectively.

We remark that the transition arises from phase-locking, which is a quasilocal, single-particle phenomenon, and not from coupling some local degree of freedom to a collective (mean-field) excitation.
This consideration sets the present setup apart from other examples~\cite{AA2002,BLS2004} of spontaneous onset of an order parameter in a many-body system.

\subsection{Analytical interpolation with displaced thermal states}
\label{ssec:interpolation}

To provide an analytical description of the nonequilibrium phase transition presented above, we derive now an effective dynamics for the field expectation value $\alpha$.
The effective dynamics that we construct is such that $\alpha$ converges to $\alpha = 0$ in the normal phase, and to $|\alpha| > 0$ in the coherent phase.
Focusing on the parameter region around the critical point, $\alpha$ performs an overdamped motion on a energy potential $U(\alpha) \simeq c_{2} |\alpha|^{2} + c_{4} |\alpha|^{4}$.
Our construction offers an effective description of the asymptotic dynamics of the system in terms of its collective variables.
In this sense, $\alpha$ plays the role of the order parameter of the dynamical phase transition.
In analogy to Landau theory, where the minimization of the free energy gives the equilibrium phase, the fixed point of the effective dynamics gives the steady state of the nonequilibrium system.

To construct the effective dynamics, we make the ansatz that, close to the steady state, the density matrix $\rho$ depends only on $\alpha$ and can be approximated by the form
\begin{equation}\label{eq:ansatz}
\rho(\alpha) \equiv \hat{\cal D}(\alpha) \bar{\rho} (\bar{n}) \hat{\cal D}(-\alpha)~,
\end{equation}
where the action of the Glauber operator $\hat{\cal D}$ is $\hat{\cal D}(-\alpha) \hat{a} \hat{\cal D}(\alpha) = \hat{a} + \alpha$, and $\bar{\rho}$ is the thermal density matrix with average density $\bar{n}$.
The total photonic density $n = \bar{n} + |\alpha|^{2}$ features an incoherent and a coherent contribution (that correspond to the thermal and condensed fractions, if the same formalism is adapted to describe ultracold bosonic matter).
Particular cases of this ansatz include the coherent state, where $n = |\alpha|^{2}$, and the thermal state where $\alpha = 0$ and the density matrix is diagonal.

The time-evolution $\partial_{t} n(t) = - 2 (\kappa - \Upsilon) n(t) + 2 \Upsilon$ of the total density is determined by the incoherent rates of injection and leakage of photons and gives $n_{\infty} = \Upsilon / (\kappa - \Upsilon)$ in the steady state.
In this respect, the incoherent part of the time-evolution plays a role analogous to the chemical potential in a system of massive bosons at equilibrium.
The fact that the total number of particles does not depend on the driving strength $\Gamma$ suggests that we may define the effective dynamics (evolving in time $\tau$) in such a way that the thermal state is always the instantaneous solution of the EOM and $\partial_{\tau} \bar{n} = 0$ (see details in App.~\ref{app:ansatz}).
In the proximity of the critical point, where the time-evolution of the order parameter $\alpha$ is expected to be much slower than any other macroscopic degree of freedom of the system, this is certainly a good approximation.
In the remaining part of the phase diagram, this approximation can be seen as a coarse graining of the dynamics.
Using the value $\bar{n} = \Upsilon / (\kappa - \Upsilon + |\alpha|^{2} \Gamma' )$, which fulfills the ansatz in the coarse-grained approximation (see App.~\ref{app:ansatz}), the effective equation of motion for the order parameter reads
\begin{equation}\label{eq:effdynamics}
\partial_{\tau} |\alpha|^{2} = -2 (\kappa - \Upsilon) \left [ |\alpha|^{2} + \frac{\Upsilon}{\kappa - \Upsilon + |\alpha|^{2} \Gamma'} - \frac{\Upsilon}{\kappa - \Upsilon} \right ]~.
\end{equation}
The stability of the system (meaning that the coherent density does not increase arbitrarily) is granted by $\kappa > \Upsilon$ because i) the prefactor multiplying the square brackets is negative and ii) the second term in the square brackets vanishes as $|\alpha|^{2} \to \infty$, with the sum of the two remaining terms positive.
It is interesting to see that, while the stability of the system for $\Gamma' = 0$ is obvious, for $\Gamma' > 0$ the effect is nonperturbative, as it stems from a term $\propto |\alpha|^{-2}$ and cannot be obtained from a generic expansion in the Landau form.
The order parameter around the critical point follows the overdamped equation $\frac{1}{m} \partial_{\tau}^{2} |\alpha| = -\partial_{\tau} |\alpha| - \partial U / \partial |\alpha|$, for $m \to \infty$, with the effective energy
\begin{equation}\label{eq:effpotential}
U = (\kappa - \Upsilon) \frac{|\alpha|^{2}}{2} + \Gamma' \left ( \frac{|\alpha|^{4}}{4} - n_{\infty} \frac{|\alpha|^{2}}{2} \right )~.
\end{equation}
The solution for the steady states of the effective dynamics (which coincide by construction with the minima of the effective potential $U$) then reads $|\alpha|^{2} = 0$ for $\Gamma' < \Gamma_{\rm crit}'$ and $|\alpha|^{2} = n_{\infty}(1 - \Gamma_{\rm crit}' / \Gamma')$ for $\Gamma' > \Gamma_{\rm crit}'$, with $\Gamma_{\rm crit}' = (\kappa - \Upsilon)^{2}/\Upsilon$.
It is interesting to note that, for any $\kappa$, one can drive the system strong enough to overcome thermalization and enter the phase where the transition takes place.
Eq.~(\ref{eq:effdynamics}) is applicable to all values of the driving $\Gamma'$, features the correct results in the limits $\Gamma \to 0$ and $\Gamma \to \infty$,  and correctly captures the existence of a nonequilibrium phase transition in the model.
The profile of $|\alpha|^{2}$ obtained by the numerical solution in the close vicinity $\delta \Gamma'/\kappa \lesssim 0.1$ of $\Gamma_{\rm crit}'$, though, is more complex than the linear slope which follows from Eq.~(\ref{eq:effpotential}).

\section{Conclusions and outlook}
\label{sec:conclusions}

In conclusion, we have devised a scheme to implement a driven-dissipative dynamics for photons in an OM setup.
The mechanical elements offer both a way to drive the photons between cavity modes, and a natural reservoir where energy can be dissipated.
Due to the huge energy difference of the photonic and the phononic modes, the engineered effective dynamics for the photons is number-conserving.
The basic block of our scheme is an effective $\Lambda$-system for photons, which generates a dark state consisting of the symmetric superposition of two photonic modes.
Extension of the scheme to an array of membranes is shown to generate a state with long-range coherence, where the phase-locking is achieved throughout the array.
The interplay of the engineered driven-dissipative dynamics with the losses of photons out of the cavity and incoherent driving of the photonic modes results in a dynamical phase transition.
We provide a compact analytical description of such transition, in terms of an effective potential that is minimized in the steady state.

We have focused on an implementation based on micromechanical membranes in an optical cavity, where the photonic modes are delimited by the membranes and the mirrors of a Fabry-P\'erot resonator.
This kind of setup offers extremely low optical losses and large tunability in the position of the mechanical oscillators.
It is an ideal candidate to test the building block of our setup, which consists of two membranes in a cavity.
However, other OM implementations, such as arrays of nanotoroids or OM crystals, provide a better scalability, and might be more suited to investigate the onset of long-range coherence in the array.
The scheme that we devised is generic and could be extended to such implementations as well.

OM setups are complementary to other quantum simulation frameworks, such as cold atoms or circuit-QED, as to which kind of many-body phenomena can be realized.
The possibility to manipulate the energy spectrum by geometrically displacing the mechanical oscillators, and the naturally arising coupling to well-controlled phononic reservoirs, make OM setups ideally suited to implement nonequilibrium models.
In this work we considered the time-evolution and stationary state of a many-body model, in the presence of continuous flow of particles through the system.
The existence of a dynamical phase transition, which is not present in the number-conserving dynamics, suggests that it is interesting to study other many-body models under such conditions as well.
Finally, once strong-coupling of photons and phonons is available in OM devices, the effect of dissipation on well-known equilibrium quantum phase transitions can also be investigated along these lines.

\acknowledgments We acknowledge the European Union through IP AQUTE, the Austrian Science Fund (FWF)
through SFB FOQUS,  the START grants Y 591-N16 (P.R.) and Y 581-N16 (S.D.) and the Institute for Quantum Information.
M.L. acknowledges support by NSF, CUA, and the Packard Foundation.
Free Software (www.gnu.org, www.python.org) was used in this work.

\appendix

\section{Localized modes in the array}
\label{app:newmodes}

In this appendix we substantiate the existence of the two families of modes $\hat{a}_{\ell}$ and $\hat{c}_{\ell}$ discussed in Sec.~\ref{sec:membranesarray}.
The procedure that we use relies on the possibility to define decaying modes $\hat{a}$ for light between leaking mirrors, as detailed in Ref.~\cite{VogelWelschBook}.
The existence of hopping processes between the internal modes of a cavity and \emph{all} the external modes of the electromagnetic spectrum, even at different frequencies, is commonly used to describe leaking cavities from the quantum noise perspective~\cite{CarmichaelBook,GardinerZollerBook}.
In our case, with the hopping term we model the existence of nonzero matrix element between decaying modes bounded by membranes and mirrors, with different energy.
In the absence of external driving, these transitions are of course only virtual and the real transport through the array takes place via double-tunneling events of amplitude $J^{2}$.

To discuss the form of the modes, we use a single-particle picture with Hamiltonian
\begin{equation}
\hat{H} = \left ( \begin{array}{cccc}
-1 / 2 & J & 0 & J \\
J & +1 / 2 & J & 0 \\
0 & J & - 1 / 2 & J \\
J & 0 & J & +1 / 2
\end{array} \right )~,
\end{equation}
which describes an array of double wells, where the energy difference $\delta\omega$ between the wells has been taken equal to $1$, and $J$ is the hopping between the wells in units of $\delta\omega$.
The basis in which the Hamiltonian is written supports the definition of the modes $\hat{a}_{\ell}$ (with energy $-1/2$) and $\hat{c}_{\ell}$ (with energy $+1/2$).
Periodic boundary conditions are used in the Hamiltonian matrix.  Now let us consider the transformation defined by the matrix
\begin{equation}
M = \left ( \begin{array}{cccc}
-1 & J & 0 & J \\
J & 1 & J & 0 \\
0 & J & -1 & J \\
J & 0 & J & 1
\end{array} \right )~,
\end{equation}
such that
$ M \, M^{\rm T} = \mathbb{1} + {\cal O}[J^{2}]$ and
\begin{equation}
M \, \hat{H} \, M^{-1}
= \left ( \begin{array}{cccc}
-1/2 & 0 & 0 & 0 \\
0 & 1/2 & 0 & 1 \\
0 & 0 & -1/2 & 0 \\
0 & 0 & 0 & 1/2
\end{array} \right ) + {\cal O}[J^{2}]~.
\end{equation}
That is, we have presented a transformation $M$ that, to first order in $J / \delta\omega$, is orthogonal (i.e.~implements a canonical basis transformation from independent bosonic modes to independent bosonic modes) and diagonalizes the hopping part of the Hamiltonian.
In other words, the dominant part of the hopping is included into the definition of the new basis states.
The new basis states support the definition of the modes $\hat{a}_{\ell}$ (with energy $\simeq -1/2$) and $\hat{c}_{\ell}$ (with energy $\simeq 1/2$).
The first correction includes hopping between next-nearest neighbors, that is between $\hat{a}_{\ell}$ and $\hat{a}_{\ell \pm 1}$ or between $\hat{c}_{\ell}$ and $\hat{c}_{\ell \pm 1}$.

\section{Separability of the steady state}
\label{app:separability}

\begin{figure}
\includegraphics[width=\linewidth]{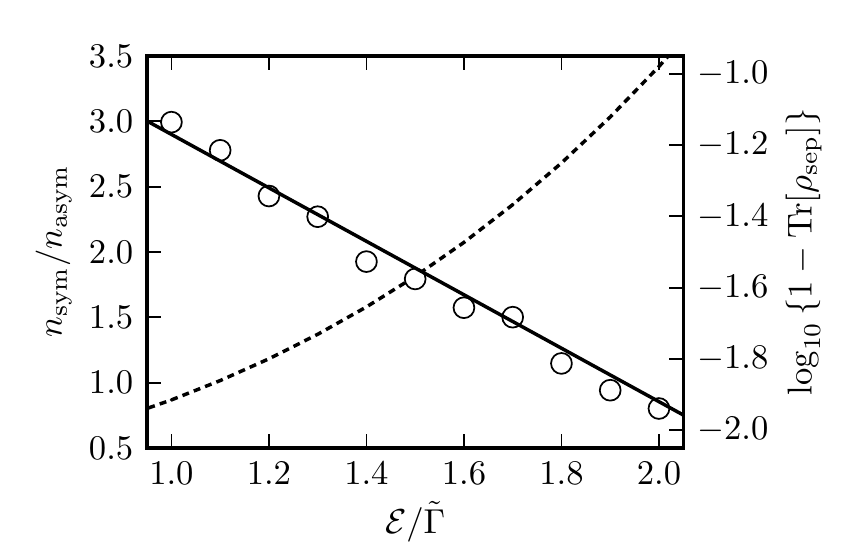}
\caption{\label{fig:increasedriving}
Properties of the steady state as the strength ${\cal E}$ of the OM driving is increased.
The ratio between the number of particles in the symmetric and antisymmetric mode (dashed line, left axis) increases, while the trace of the separable approximation $\rho_{\rm sep}$ of Eq.~(\ref{eq:separableapproximation}) to the full density matrix (empty circles, right axis) approaches unity exponentially (solid line).
We use $\varepsilon_{1} = 0.2 \tilde{\Gamma}$, $\Delta_{1} = \tilde{J} = 0.2 \tilde{\Gamma}$, $\kappa = 0.1 \tilde{\Gamma}$, and $\Delta = 2.0 \tilde{\Gamma}$.
The state has been evolved to $t_{\rm max} \tilde{\Gamma} = 10.0$.
The truncation of the Fock space is $n_{\rm max} = 5$ for the modes $\hat{a}_{\ell}$ and $n_{\rm max} = 1$ for the mode $\hat{c}$, which is negigibly populated.  }
\end{figure}

In this appendix we discuss in more detail the effect of switching on the OM coupling in the two-membrane system introduced in Sec.~\ref{sec:twomembranes}.
As we show in Fig.~\ref{fig:increasedriving}, the ratio betwen the population in the symmetric mode and the population in the antisymmetric mode increases monotonically with the strength of the driving.
One has to keep in mind, though, that the strength cannot be increased indefinitely because the model that we use here adopts the rotating-wave approximation, which is not justified for arbitrary large drivings.
The symmetric mode in the steady state is not only increasingly populated, but also increasingly disentangled from the remaining part of the system, as the driving is increased.
To demonstrate this, we produce an ansatz for density matrix in the steady state, in the form of a separable state $\rho_{\rm sep}$ plus some remainder $\sigma_{\rm ent}$, and we show that the contribution of $\rho_{\rm sep}$ to the total trace converges to unity exponentially.
More precisely, we first diagonalize the steady-state density matrix as $\rho = \sum_{\lambda} p_{\lambda} |\Psi_{\lambda} \rangle \langle \Psi_{\lambda}|$, where $p_{\lambda}$ is a set of probability values that sums to unity and $|\Psi_{\lambda}\rangle$ is a pure state.
Each state $|\Psi_{\lambda}\rangle$ can be expanded in the tensor basis of the modes as $|\Psi_{\lambda}\rangle = \prod_{n_{\rm s}n_{\rm a}n_{3}} \Psi_{\lambda;n_{\rm s}n_{\rm a}n_{3}} |n_{\rm s}\rangle|n_{\rm a}\rangle |n_{3}\rangle$, where $n_{\rm s}$, $n_{\rm a}$, $n_{3}$ are the indices of the Fock states in the symmetric, antisymmetric, and $\hat{c}$ modes, respectively.
We take $n_{3} = 0$, corresponding to a negligibly populated excited state, and retain only the terms in the sum with the Fock occupation $\bar{n}_{\rm s}^{(\lambda)}$ that maximizes $\sum_{n_{\rm a}}|\Psi_{\lambda;\bar{n}_{\rm s}^{(\lambda)},n_{\rm a},0}|^{2}$.
The resulting expression
\begin{eqnarray}\label{eq:separableapproximation}
\rho_{\rm sep} & = & \sum_{\lambda} p_{\lambda} | \bar{n}_{\rm s}^{(\lambda)}
\rangle \langle \bar{n}_{\rm s}^{(\lambda)} | \otimes \sum_{n_{\rm a}n_{\rm a}'}
\Psi_{\lambda;\bar{n}_{\rm s}^{(\lambda)},n_{\rm a},0}
\Psi_{\lambda;\bar{n}_{\rm s}^{(\lambda)},n_{\rm a}',0}^{\ast} \nonumber \\
& & | n_{\rm a} \rangle \langle n_{\rm a}' | \otimes |n_{3} = 0 \rangle \langle
n_{3} = 0|
\end{eqnarray}
has the form of a separable state by construction, but non-unitary trace $\sum_{\lambda} \sum_{n_{\rm a}} p_{\lambda} |\Psi_{\lambda;\bar{n}_{\rm s}^{(\lambda)},n_{\rm a},0}|^{2}$, and the full density matrix can be trivially decomposed as $\rho = \rho_{\rm sep} + \sigma_{\rm ent}$.
We see that, as the driving strength increases, the trace of $\rho_{\rm sep}$ saturates the unitary total trace exponentially, meaning that the ansatz becomes numerically exact for strong driving.
The separability of the steady state should be seen here as analogous to the purity of the steady state as stated in Eq.~(\ref{eq:purity}).
Namely, the reduced density matrix $\rho_{\rm s} = {\rm Tr}_{n_{\rm a},n_{3}} \langle n_{\rm a} | \langle n_{3} | \rho | n_{\rm a} \rangle | n_{3} \rangle$ of one mode is not a pure state both because of the entanglement between the modes, and because of the local coupling to the external reservoirs, which is effectively switched off in the number conserving models of driven-dissipative systems~\cite{DMKKBZ2008}.
In other words, the entanglement entropy, i.e.~the von Neumann entropy of the reduced density matrix, is a good measure of entanglement for states that are pure, while we consider a system which is continuously subjected to pumping and losses.
Separability is, in this respect, the closest concept that we may apply here to characterize the fact that the combined process of driving and dissipation does not entangle the dark state of the system (the symmetric mode) to the other (auxiliary) modes.
One may also expect that the steady state is actually a product state, i.e.~a separable state where only one element in the sum over $\lambda$ is finite.
We conjecture that this expectation is fulfilled in the limit where the population of the states is very large, when the pumped antisymmetric state is substantially insensitive to the losses towards the symmetric state.
Intuitively, a larger intensity in the symmetric mode is correlated to a smaller intensity in the antisymmetric mode, because the coupling between the two modes is not negligible with respect to their internal dynamics.
In the limit of large populations, however, the antisymmetric state can be effectively seen as a bath that injects particles into the symmetric state, and hence the total system factors.

\section{Interpolation ansatz}
\label{app:ansatz}

To validate the ansatz (\ref{eq:ansatz}), one must make sure that the displaced thermal state is a solution to the equation ${\cal L}[\rho(\alpha)] = 0$ for the steady state.
Using the properties of the Glauber operators in the definition of $\rho(\alpha)$, we first compute the Liouvillian ${\cal L}_{\rm d}[\alpha; \bar{\rho}] = \hat{\cal D}(-\alpha) {\cal L}[\rho(\alpha)] \hat{\cal D}(\alpha)$ in the displaced frame, which acts on the diagonal density matrix $\bar{\rho}$ and where $\alpha$ is a parameter.
The equation ${\cal L}_{\rm d}[\alpha; \bar{\rho}] = 0$ for the steady state is then treated within the coarse-grained approximation mentioned above.
More precisely, in the equation of motion we neglect linear and cubic terms, which perturb the diagonal thermal state but do not substantially change the approximate steady-state manifold defined by the ansatz.
The dominant part of the coherent fraction is contained in the displacement produced by the Glauber operators, while we neglect the quantum fluctuations of the coherent fraction induced by linear and cubic terms.
Such quantum fluctuations are expected to be much faster than the slow dynamics of the order parameter $\alpha$, and in this sense this approximation can be understood as a coarse-graining of the effective dynamics.
The quartic terms vanish identically on the diagonal density matrix, while the remaining quadratic terms read
\begin{eqnarray}\label{eq:displacedliouv}
{\cal L}_{\rm d}[\alpha; \bar{\rho}] & \simeq & \tilde{\kappa}(\bar{n}) [2 \hat{a} \bar{\rho} \hat{a}^{\dag} - \{ \hat{n}, \bar{\rho} \} ] \vphantom{\tilde{\Upsilon}} \nonumber \\
& + & \tilde{\Upsilon}(\bar{n}) [2 \hat{a}^{\dag} \bar{\rho} \hat{a} - \{ \hat{a} \hat{a}^{\dag}, \bar{\rho} \} ]~.
\end{eqnarray}
This form consists of effective dissipation of width $\tilde{\kappa}(\bar{n}) = \kappa + \bar{n} \frac{3}{13}\Gamma' + (|\alpha|^{2} + \frac{3}{13})\Gamma'$ and incoherent pumping of strength $\tilde{\Upsilon}(\bar{n}) = \Upsilon + \bar{n} \frac{3}{13} \Gamma'$, with $\Gamma' = \frac{156}{3} \Gamma$.
The effect of the engineered dissipation $\propto \Gamma$ is thus to renormalize the loss and pumping coefficients in the displaced frame.
The right-hand side of Eq.~(\ref{eq:displacedliouv}) vanishes if the thermal contribution to the density fulfills $\bar{n} = \tilde{\Upsilon}(\bar{n}) / [\tilde{\kappa}(\bar{n}) - \tilde{\Upsilon}(\bar{n})]$, the solution of which gives
\begin{equation}
\bar{n} = \frac{\Upsilon}{\kappa - \Upsilon + |\alpha|^{2} \Gamma'}~.
\end{equation}
The rescaled driving strength $\Gamma'$ includes model-dependent parameters so that the formula for $\bar{n}$ is simple.
The large factor between $\Gamma'$ and $\Gamma$ is due to the complicated form of the mean-field Liouvillian, where many nonlocal terms add up.
We verified that a different model Liouvillian, which describes onsite driving of photons between bands in an array of equally spaced membranes, reduces as well to the present form, with different but equivalent functions $\tilde{\kappa}$ and $\tilde{\Upsilon}$.

\section{Coefficients of the Gutzwiller approximation}
\label{app:coefficients}

We derived the following convenient formula, which gives the mean-field decoupling for a Liouvillian in a non-diagonal Lindblad form, coupling up to $M$ sites together
\onecolumngrid
\begin{eqnarray}
& & \sum_{\ell'} {\rm Tr}_{\neq \ell} W_{\ell'} \left [ 2 \left ( \prod_{m=1}^{M} \hat{A}_{\ell'+\alpha_{m}}^{(m)} \right ) \rho \left ( \prod_{m=1}^{M} \hat{B}_{\ell'+\alpha_{m}}^{(m)} \right ) - \left \lbrace \left ( \prod_{m=1}^{M} \hat{B}_{\ell'+\alpha_{m}}^{(m)} \right ) \left ( \prod_{m=1}^{M} \hat{A}_{\ell'+\alpha_{m}}^{(m)} \right ) , \rho \right \rbrace \right ] \nonumber \\
& & = \sum_{m=1}^{M} W_{\ell-\alpha_{m}} \left ( \prod_{m' \neq m} \langle \hat{B}^{(m')} \hat{A}^{(m')} \rangle_{\ell + \alpha_{m'} - \alpha_{m}} \right ) \times \left [ 2 \hat{A}^{(m)}_{\ell} \rho_{\ell} \hat{B}^{(m)}_{\ell} - \{ \hat{B}^{(m)}_{\ell} \hat{A}^{(m)}_{\ell} , \rho \} \right ]~,
\end{eqnarray}
with the definition of the local average $\langle \hat{A} \rangle_{\ell} = {\rm Tr}_{\ell} \left [ \hat{A}_{\ell} \rho_{\ell} \right ]$.
The coefficients of the mean-field Liouvillian in Eq.~(\ref{eq:mfliouv}) read
\begin{equation}
L^{(1)} =
\left ( \begin{array}{cccc}
0
&
\begin{array}{l}
2 \{ 2 \langle \hat{a}^{\dag} \rangle [1 + \chi + \chi^{2} + \\
-\chi ( 1 + \chi) \langle \hat{a}^{\dag} \hat{a} \rangle] + \\
+ (4 + 7 \chi + 4 \chi^{2}) \langle \hat{a}^{\dag} \hat{a}^{\dag} \hat{a} \rangle  \}
\end{array}
&
0
&
0
\\
\begin{array}{l}
2 \{ 2 \langle \hat{a} \rangle [1 + \chi + \chi^{2} + \\
-\chi ( 1 + \chi) \langle \hat{a}^{\dag} \hat{a} \rangle] + \\
+ (4 + 7 \chi + 4 \chi^{2}) \langle \hat{a}^{\dag} \hat{a} \hat{a} \rangle  \}
\end{array}
&
\begin{array}{l}
4 [ \langle \hat{a}^{\dag} \hat{a} \rangle + \chi ( 1 + \chi) \times \\
\times ( \langle \hat{a}^{\dag} \rangle \langle \hat{a} \rangle + \langle \hat{a}^{\dag} \hat{a} \rangle )  ]
\end{array}
&
\begin{array}{l}
-2 ( \chi \langle \hat{a} \rangle^{2} + \\
+ 2 (1 + \chi ) \langle \hat{a}^{2} \rangle)
\end{array}
&
-4 (1 + 2 \chi + 2 \chi^{2}) \langle \hat{a} \rangle
\\
0
&
\begin{array}{l}
-2 ( \chi \langle \hat{a}^{\dag} \rangle^{2} + \\
+ 2 (1 + \chi) \langle \hat{a}^{\dag 2} \rangle )
\end{array}
&
\begin{array}{l}
4 (1 + \chi + \chi^{2}) \times \\
\times ( \langle \hat{a}^{\dag} \hat{a} \rangle + 1 )
\end{array}
&
2 ( 2 + 3 \chi) \langle \hat{a}^{\dag} \rangle
\\
0
&
-4 (1 + 2 \chi + 2 \chi^{2}) \langle \hat{a}^{\dag} \rangle
&
2 (2 + 3 \chi) \langle \hat{a} \rangle
&
4(1 + 2 \chi + 2 \chi^{2}) 
\end{array} \right ) ~,
\end{equation}
\begin{equation}
L^{(2)} =
\left ( \begin{array}{cccc}
0
&
\begin{array}{l}
- 2 \{ \langle \hat{a}^{\dag} \rangle [ (2 + 3 \chi) + \\
- 2 \chi ( 1 + \chi) \langle \hat{a}^{\dag} \hat{a} \rangle] + \\
+ (4 + 7 \chi + 4 \chi^{2}) \langle \hat{a}^{\dag} \hat{a}^{\dag} \hat{a} \rangle  \}
\end{array}
&
0
&
0
\\
\begin{array}{l}
- 2 \{ \langle \hat{a} \rangle [(2 + 3 \chi) + \\
- 2 \chi ( 1 + \chi) \langle \hat{a}^{\dag} \hat{a} \rangle] + \\
+ (4 + 7 \chi + 4 \chi^{2}) \langle \hat{a}^{\dag} \hat{a} \hat{a} \rangle  \}
\end{array}
&
4 (1 + \chi + \chi^{2}) \langle \hat{a}^{\dag} \hat{a} \rangle
&
\begin{array}{l}
-2 ( \chi \langle \hat{a} \rangle^{2} + \\
+ 2 (1 + \chi ) \langle \hat{a}^{2} \rangle)
\end{array}
&
2 (2 + 3 \chi) \langle \hat{a} \rangle
\\
0
&
\begin{array}{l}
-2 [ \chi \langle \hat{a}^{\dag} \rangle^{2} + \\
+ 2( 1 + \chi ) \langle \hat{a}^{\dag 2} \rangle]
\end{array}
&
\begin{array}{l}
4 [\chi ( 1 + \chi) \langle \hat{a}^{\dag} \rangle \langle \hat{a} \rangle + \\
+(1 + \chi + \chi^{2}) \times \\
\times (\langle \hat{a}^{\dag} \hat{a} \rangle + 1)]
\end{array}
&
-4 ( 1 + 2 \chi + 2 \chi^{2}) \langle \hat{a}^{\dag} \rangle
\\
0
&
2 ( 2 + 3 \chi) \langle \hat{a}^{\dag} \rangle
&
-4 (1 + 2  \chi + 2 \chi^{2} ) \langle \hat{a} \rangle
&
4(1 + 2 \chi + 2 \chi^{2})
\end{array} \right ) ~.
\end{equation}

\twocolumngrid

\end{document}